\documentclass[aps, prl,reprint, amsmath,amssymb, superscriptaddress]{revtex4-1}

\usepackage{graphicx}
\usepackage{xcolor,color}

\begin{document}

\title{Role of matrix elements in the time-resolved photoemission signal}

\author{F.\,Boschini}
\email[]{boschini@phas.ubc.ca}
\affiliation{Department of Physics $\&$ Astronomy, University of British Columbia, Vancouver, BC V6T 1Z1, Canada}
\affiliation{Quantum Matter Institute, University of British Columbia, Vancouver, BC V6T 1Z4, Canada}
\author{D. Bugini}
\affiliation{Dipartimento di Fisica, Politecnico di Milano, 20133 Milano, Italy}
\affiliation{Center for Nano Science and Technology@PoliMi, Istituto Italiano di Tecnologia, 20133 Milano, Italy}
\author{M.\,Zonno}
\affiliation{Department of Physics $\&$ Astronomy, University of British Columbia, Vancouver, BC V6T 1Z1, Canada}
\affiliation{Quantum Matter Institute, University of British Columbia, Vancouver, BC V6T 1Z4, Canada}
\author{M.\,Michiardi}
\affiliation{Department of Physics $\&$ Astronomy, University of British Columbia, Vancouver, BC V6T 1Z1, Canada}
\affiliation{Quantum Matter Institute, University of British Columbia, Vancouver, BC V6T 1Z4, Canada}
\affiliation{Max Planck Institute for Chemical Physics of Solids, Dresden, Germany}
\author{R.\,P.\,Day}
\author{E.\,Razzoli}
\author{B.\,Zwartsenberg}
\affiliation{Department of Physics $\&$ Astronomy, University of British Columbia, Vancouver, BC V6T 1Z1, Canada}
\affiliation{Quantum Matter Institute, University of British Columbia, Vancouver, BC V6T 1Z4, Canada}
\author{E.\,H.\,da Silva Neto}
\affiliation{Department of Physics $\&$ Astronomy, University of British Columbia, Vancouver, BC V6T 1Z1, Canada}
\affiliation{Quantum Matter Institute, University of British Columbia, Vancouver, BC V6T 1Z4, Canada}
\affiliation{Max Planck Institute for Solid State Research, Heisenbergstrasse 1, D-70569 Stuttgard, Germany}
\affiliation{Department of Physics, University of California, Davis, CA 95616, USA}
\author{S. dal Conte}
\affiliation{Dipartimento di Fisica, Politecnico di Milano, 20133 Milano, Italy}
\author{S. K. Kushwaha}
\affiliation{Department of Chemistry, Princeton University, Princeton, New Jersey 08544, USA}
\affiliation{National High Magnetic Field Laboratory, LANL, Los Alamos, New Mexico 87504, USA}
\author{R. J. Cava}
\affiliation{Department of Chemistry, Princeton University, Princeton, New Jersey 08544, USA}
\author{S.\,Zhdanovich}
\author{A.\,K.\,Mills}
\author{G.\,Levy}
\affiliation{Department of Physics $\&$ Astronomy, University of British Columbia, Vancouver, BC V6T 1Z1, Canada}
\affiliation{Quantum Matter Institute, University of British Columbia, Vancouver, BC V6T 1Z4, Canada}
\author{E. Carpene}
\affiliation{IFN-CNR, Dipartimento di Fisica, Politecnico di Milano, 20133 Milano, Italy}
\author{C. Dallera}
\affiliation{Dipartimento di Fisica, Politecnico di Milano, 20133 Milano, Italy}
\author{C.\,Giannetti}
\affiliation{Department of Mathematics and Physics, Universit\`{a} Cattolica del Sacro Cuore, Brescia, BS I-25121, Italy}
\affiliation{Interdisciplinary Laboratories for Advanced Materials Physics (ILAMP),Universit\`{a} Cattolica del Sacro Cuore, Brescia I-25121, Italy}
\author{D.\,J.\,Jones}
\affiliation{Department of Physics $\&$ Astronomy, University of British Columbia, Vancouver, BC V6T 1Z1, Canada}
\affiliation{Quantum Matter Institute, University of British Columbia, Vancouver, BC V6T 1Z4, Canada}
\author{G. Cerullo}
\affiliation{Dipartimento di Fisica, Politecnico di Milano, 20133 Milano, Italy}
\author{A.\,Damascelli}
\email[]{damascelli@physics.ubc.ca}
\affiliation{Department of Physics $\&$ Astronomy, University of British Columbia, Vancouver, BC V6T 1Z1, Canada}
\affiliation{Quantum Matter Institute, University of British Columbia, Vancouver, BC V6T 1Z4, Canada}

\begin{abstract}
Time- and angle-resolved photoemission spectroscopy accesses the ultrafast evolution of quasiparticles and many-body interactions in solid-state systems. However, the momentum- and energy-resolved transient photoemission intensity may not be unambiguously related to the intrinsic relaxation dynamics of photoexcited electrons. In fact, interpretation of the time-dependent photoemission signal can be affected by the transient evolution of both the one-electron removal spectral function as well as the photoemission dipole matrix elements.
Here we investigate the topological insulator Bi$_{1.1}$Sb$_{0.9}$Te$_2$S to demonstrate, by means of a careful probe-polarization study, the transient contribution of matrix elements to the time-resolved photoemission signal.
\end{abstract}

\pacs{78.47.J-, 74.25.Jb, 79.60.-i}

\maketitle

\section{Introduction\label{Introduction}}
The development of pump-probe techniques has provided the opportunity to extend the study of solid state systems into the time domain, garnering important insights regarding transient phenomena in addition to new perspectives on persistent challenges from equilibrium \cite{ReviewGiannetti}. Generally speaking, pump-probe techniques rely on a simple principle: a pump pulse drives the system out-of-equilibrium while a delayed probe pulse tracks intrinsic scattering properties on an ultrafast time scale. \\
The momentum information accessible from time- and angle-resolved photoemission spectroscopy (TR-ARPES) offers a significant advantage over other pump-probe techniques, as the modifications to the electronic structure and momentum-resolved relaxation dynamics of photoexcited electrons are observed directly. TR-ARPES has been used to study the transient evolution of exotic phases in condensed matter such as unconventional superconductivity \cite{Boschini2018,ScienceLanzara,BovensiepenFebased}, charge-order \cite{CollapseCDWTiSe2}, excitonic condensates \cite{StahlerExcitonic} and Floquet states \cite{GedikFloquetScience,GedikFloquetNatPhys}. 
While the technique is by now fairly well-established, interpretation and analysis of TR-ARPES has yet to take advantage of the vast quantity of information encoded in the experimental signal. Presently, it is conventional to emphasize the temporal evolution of the electronic temperature \cite{BovensiepenFebased,SterziSmB6,SterziTI,PRL2007PerfettiBi2212} or the photoemission intensity in well-defined momentum-energy regions \cite{CrepaldiWeyl,PerfettiBiTe,ShenBiSe2012,BuginiBiSe}. While in many cases this approach provides a basic understanding of some transient properties of the electronic population, a comprehensive description of the experiment is challenging.\\
To explore this further, we consider the form of the photoemission signal, for fixed energy $\omega$ and momentum $\textbf{k}$, as it is defined via Fermi's Golden Rule \cite{ReviewDamascelli}
\begin{equation} \label{EQ:1}
   I_{\text{PES}}(\omega,\textbf{k})=|M_{f,i}^{\textbf{k}}|^2 \cdot A(\omega,\textbf{k})\cdot f(\omega,\textbf{k}),
\end{equation}
where $|M_{f,i}^{\textbf{k}}|^2$ is the photoemission matrix element, $A(\omega,\textbf{k})$ the one-electron removal spectral function, and $f(\omega,\textbf{k})$ the electronic distribution function. It is important to note that Eq.\,\ref{EQ:1} is only strictly valid at equilibrium and may not accurately describe the effect of the pump's electric field on the ground-state Hamiltonian \cite{FaustiBeyondEnergyRes}. However, when pump and probe beams are not synchronous, Eq.\,\ref{EQ:1} can be taken to approximate the transient photoemission signal, \emph{i.e.} extended in the time ($\tau$) domain. 
The temporal evolution of $f(\omega,\textbf{k},\tau)$ describes the intrinsic relaxation processes for fixed $\omega$ and $\textbf{k}$. In addition to $f(\omega,\textbf{k},\tau)$, much emphasis has also been placed on the evolution of $A(\omega,\textbf{k},\tau)$, which encodes information regarding the bare electronic dispersion \cite{ShenBiSe2014,ScienceShenTRxrd} and many-body interactions. Such dynamical analysis of $A(\omega,\textbf{k},\tau)$ has been applied successfully to, for example, ultrafast metal-insulator transitions \cite{PerfettiTaS2,StahlerExcitonic,CollapseCDWTiSe2} as well as the quenching of phase coherence in superconducting condensates \cite{Boschini2018}.  
To date however, the possible role of the matrix element term, $|M_{f,i}^{\textbf{k}}(\tau)|^2$, has been neglected in the analysis of TR-ARPES experiments. Derived from the dipole-operator matrix elements connecting the initial ($i$) and final ($f$) state of the photoemitted electron, $|M_{f,i}^{\textbf{k}}(\tau)|^2$ is often argued to depend solely on the experimental geometry and orbital symmetry of the initial states. Wherein this assumption is valid, and the excitation is not anticipated to influence the orbital symmetry, the emphasis on $A(\omega,\textbf{k},\tau)$ and $f(\omega,\textbf{k},\tau)$ is perfectly reasonable. For multi-orbital systems however, the matrix element becomes a more complex object, susceptible to photoelectron interference effects which can modulate the relative photoemission intensity by orders of magnitude \cite{DamascelliBiSe2013,DamascelliBiSe2014,Ast-graphene-interference,Chiang-graphene}. This term is then sensitive to modifications of both the electronic and lattice structures, complicating the interpretation of relaxation dynamics immensely. The possibility that the temporal evolution of $A(\omega,\textbf{k},\tau)$, $|M_{f,i}^{\textbf{k}}(\tau)|^2$ and $f(\omega,\textbf{k},\tau)$ are intertwined raises important questions regarding the degree of confidence with which the ultrafast evolution of ARPES intensity can be associated with electronic relaxation dynamics alone.\\
To illustrate this point, Figure\,\ref{Fig1}(a) models how a simulated transient modification of the matrix element $|M(\tau)|^2$ (blue line, modeled as a single-exponential function with a decay time of 4\,ps convoluted with a 0.25\,ps Gaussian function) may affect the extracted electron relaxation time. The intrinsic electron relaxation time, described by the evolution of the electronic distribution $f(\tau)$ for a well-defined momentum-energy region, is taken for this example to be 7\,ps (black dashed line, convoluted with a 0.6\,ps Gaussian function). When fitting the modeled transient photoemission intensity (red line, proportional to $f(\tau)\cdot|M(\tau)|^2$) with a single-exponential, the transient increase of the matrix element factor results in the erroneous extraction of a 5.8\,ps relaxation time. This corresponds to a nearly 20$\%$ error in the extracted relaxation time, emphasizing the complications which may arise through negligence of the matrix element factor's dynamical response.\\
Such a scenario is not merely hypothetical: we demonstrate here a real example of this complicated co-evolution, observed in the relaxation dynamics of the topological insulator (TI) Bi$_{1.1}$Sb$_{0.9}$Te$_2$S (BSTS) \cite{BSTSNatComm}. We observe that the pump pulse coherently excites phonon modes which affect both the lattice and electronic structures. This results in a modification of the angular intensity distribution (\emph{i.e.} matrix element) as well as the band dispersion of the topological surface state (TSS). While such a study is in principle possible in a variety of materials, TIs such as BSTS are ideally-suited to our purposes owing to their strong response to an IR pump in terms of transient occupation of the TSS above the Fermi energy ($E_F$) \cite{HoferTHZpumpTI,WarpingBiTeRader2014, BuginiBiSe,ShenBiSe2012,ShenBiSe2PPE}, and susceptibility to the optical excitation of phonon modes \cite{ShenBiSe2014}. BSTS is preferred in this particular case to, for example, Bi$_{2}$Se$_{3}$/Te$_{3}$ due to the high chemical stability of the surface and the large bulk energy gap, which facilitates consideration of the TSS in isolation from the bulk-conduction states. \\
Pump-probe TR-ARPES experiments were conducted using a 1.55\,eV pump and 6.2\,eV probe beam, with photoemitted electrons collected via a hemispherical electron analyzer (SPECS Phoibos 150 -- overall momentum, energy and temporal resolutions are $<$0.003\,$\text{\AA}^{-1}$, 19 meV and 250 fs, respectively). To corroborate our findings, we conducted similar measurements via high-temporal resolution TR-reflectivity, confirming the presence of photoinduced coherent A$_{1g}$ optical and acoustic phonon modes. To then extract the contribution of $|M(\tau)|^2$ to the observed dynamics, a polarization-dependent TR-ARPES study was conducted, elucidating the essential role of this term in the apparent relaxation dynamics of the electronic spectral function.

\section{Results\label{Results}}
\begin{figure} 
\centering
\includegraphics[scale=1]{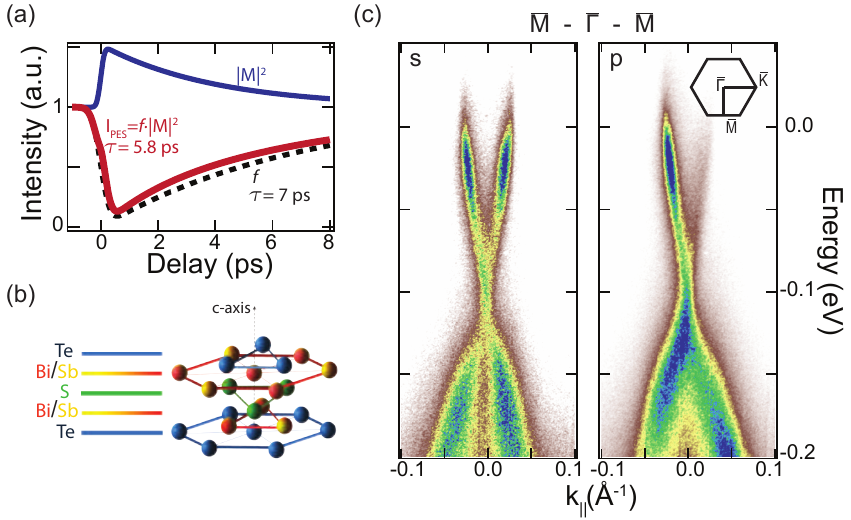}
\caption[Fig1]{(a) Pictorial sketch of how a transient modification of the matrix element $|M|^2$ (blue line, decay time 4\,ps) may affect the extraction of the electron relaxation time from the evolution of the photoemission intensity. Rather than the intrinsic 7\,ps decay time associated to the dynamics of the electronic distribution $f$ (black dashed line), a single-exponential fit of the photoemission intensity (red line) gives $\approx$5.8\,ps. (b) Bi$_{1.1}$Sb$_{0.9}$Te$_2$S layered-crystal structure. (c) Static 6.2 eV ARPES maps, along $\bar{ \Gamma }$-$\bar{\text{M}}$, acquired at 6\,K with vertically (s) and horizontally (p) polarized light. The inset shows the hexagonal projected Brillouin zone with two high-symmetry directions. }
\label{Fig1} 
\end{figure}
\begin{figure*} [t]
\centering
\includegraphics[scale=1]{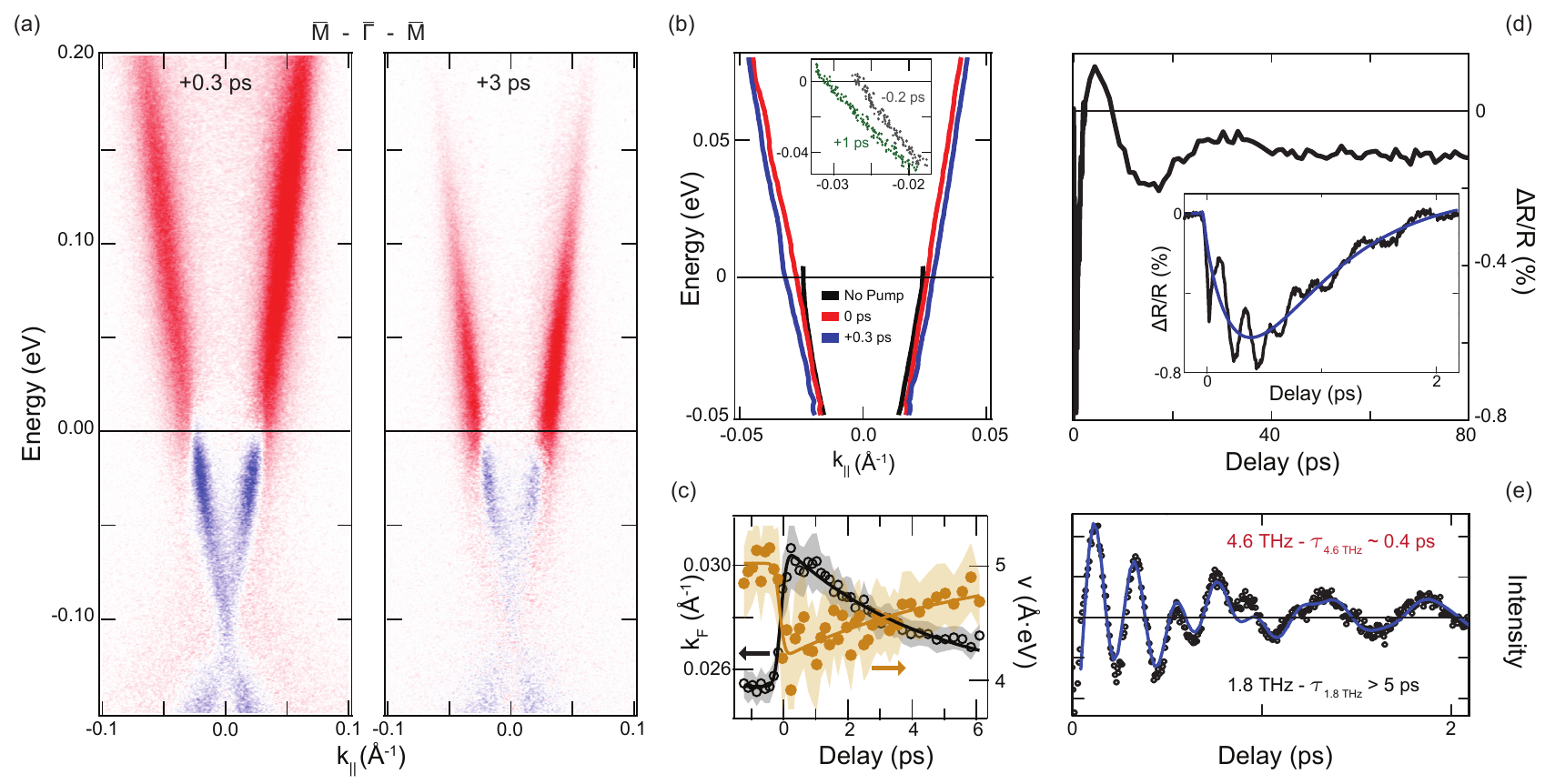}
\caption[Fig2]{(a) Differential (Pump$_{\text{ON}}$-Pump$_{\text{OFF}}$) band mapping of the TSS at +0.3 ps and +3 ps pump-probe delays, along $\bar{ \Gamma }$-$\bar{\text{M}}$, probed with s-polarized light. (b) TSS dispersion extracted by a double-Lorentzian fit of the momentum-distribution curves. The TSS dispersion changes upon the optical excitation. The inset highlights the transient modification of both the Fermi velocity and Fermi momentum by simple visual inspection. (c) Extracted Fermi velocity (v, filled orange circles) and Fermi momentum (k$_F$, open black circles) referred to the equilibrium $E_F$. Traces are fit by a phenomenological exponential decay function. The shadowed area identifies the confidence of the fitting procedure. (d) Differential reflectivity $\Delta$R($\tau$)/R trace (black curve) at 300 K, pump 2.25 eV (fluence of 3 mJ/cm$^2$) and probe 1.37 eV. The inset highlights the fast dynamics within the first 2 ps. $\Delta$R($\tau$)/R trace is fit by a phenomenological bi-exponential decay function (blue line) extracting a rise time of 0.33$\pm$0.05\,ps and subsequent relaxation time of 0.65$\pm$0.1\,ps. (e) $\Delta$R($\tau$)/R curve subtracted by the bi-exponential fitting function shown in (d). The blue line is a double damped-sinusoidal function fit.}
\label{Fig2} 
\end{figure*} 
Similar to Bi$_{2}$Se$_{3}$/Te$_{3}$, the crystal structure of BSTS forms a quintuple-layer structure with alternating layers of Te-Bi/Sb-S-Bi/Sb-Te stacked along the lattice c-axis (Fig.\,\ref{Fig1}(b)) \cite{BSTSNatComm}. Static ARPES, along $\bar{\Gamma}$-$\bar{\text{M}}$ direction and acquired with both vertically (s) and horizontally (p) polarized 6.2\,eV light, confirms the presence of a TSS, with the Dirac point located 120\,meV below $E_F$, as shown in Fig.\,\ref{Fig1}(c). Similar to other TIs, the angular distribution of ARPES intensity reflects a complex interlayer photoelectron interference effect; on the basis of orbital symmetry alone, such a pattern cannot be anticipated \cite{Cao2013,DamascelliBiSe2013}. \\
By introducing the 1.55\,eV pump excitation (s-polarized with a fluence of 40 $\mu$J/cm$^2$ throughout this work for TR-ARPES measurements) we observe not only a relative depletion/occupation of the lower/upper branches of the Dirac cone, but also a significant change to the electronic dispersion. In Fig.\,\ref{Fig2}(a), we plot the differential band-mapping of the TSS along the $\bar{\Gamma}$-$\bar{\text{M}}$ direction at 0.3\,ps and 3\,ps pump-probe delays, as probed with s-polarized light. To illustrate the pump-induced modification of the dispersion, shown in Fig.\,\ref{Fig2}(b), we plot the Fermi velocity (v) and Fermi surface area (\emph{i.e.} the Fermi momentum k$_F$) versus the pump-probe delay in Fig.\,\ref{Fig2}(c).
The expansion of the Fermi surface (referenced here to the equilibrium chemical potential) is associated with a decrease in the Fermi velocity (similar relaxation times $\approx$5\,ps, see Fig.\,\ref{Fig2}(c)), resulting in a Dirac point stable in energy over pump-probe delay - confirmed also by visual inspection of Fig.\,\ref{Fig2}(a). \\
\begin{figure*} 
\centering
\includegraphics[scale=1]{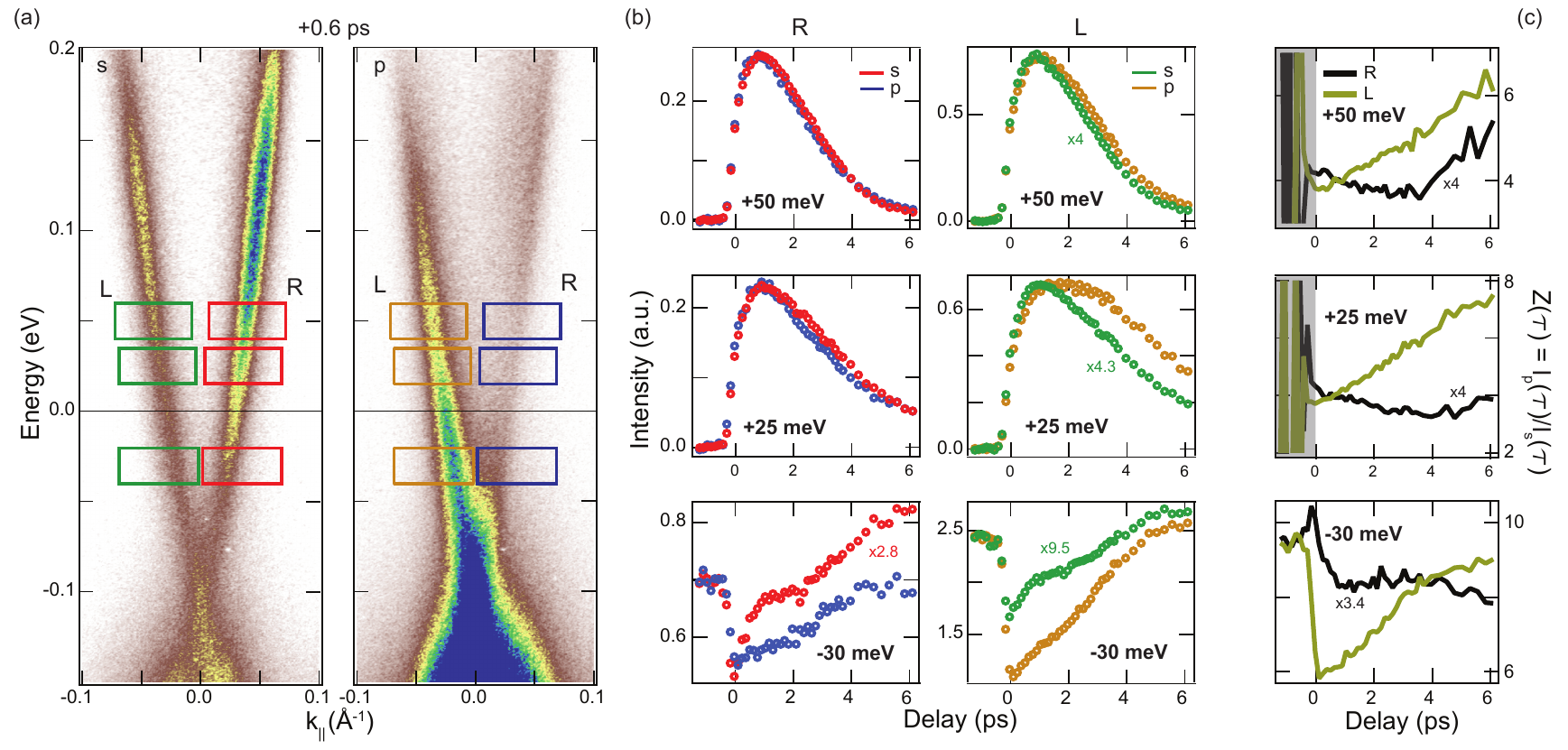}
\caption[Fig3]{(a) Out-of-equilibrium band mapping of the TSS, +0.6 ps pump-probe delay, probed with s (left) and p (right) polarizations. In each panel, the color scale is scaled to the maximum intensity of the dataset. (b) $\Delta I(\tau)$ curves centered at +50 meV, +25 meV and -30 meV resulting from the integration in colored boxes in (a). Red and blue curves: s and p polarizations, right (R) branch; green and orange curves: s and p polarizations, left (L) branches. (c) Temporal evolution of Z($\tau$) as defined in Eq.\,\ref{EQ:2}.}
\label{Fig3} 
\end{figure*}
The photoinduced modification of the dispersion of the TSS is consistent with what has been reported by Sobota et al \cite{ShenBiSe2014}, and it may be attributed to a photoinduced lattice distortion, \emph{i.e.} phonons, which affects the covalency for those states which comprise the TSS. Such phonon excitations have been reported extensively for both TIs and other materials \cite{ShenBiSe2014,ScienceShenTRxrd}. It is however possible for the pump to couple to a variety of different bosonic modes; to confirm the role of phonons here, we have performed high-resolution TR-reflectivity measurements using sub-20-fs pump and probe pulses, which are sufficiently short to impulsively excite phonons and detect them in the time domain \cite{BridaSetupMilan}. In these experiments the sample was pumped at 2.25 eV and probed with a broadband near-infrared pulse. In Fig.\,\ref{Fig2}(d-e), we demonstrate the coherent photoexcitation of two optical phonons (A$^{2}_{1g}$ 4.6\,THz and A$^{1}_{1g}$ 1.79\,THz), in addition to an acoustic phonon at 34\,GHz \cite{ShenBiSe2014,BoschiniSciRep,AcousticPhononBiSe}.
A single trace of the recorded differential reflectivity $\Delta$R($\tau$)/R at probe energy 1.37 eV \cite{noteReflectivity} is plotted in Fig.\,\ref{Fig2}(d). Within the first 2\,ps, the optical phonons are observed, while on a longer time scale the optical response is modulated by an over-damped acoustic phonon mode (34\,GHz, 0.14\,meV). Subtracting the bi-exponential fit (see inset of Fig.\,\ref{Fig2}(d)), the residual curve can be fit to damped sinusoids, from which we find that the optical modes reflect damping times of 0.4\,ps (4.6\,THz/19\,meV) and $>$5\,ps (1.79\,THz/7.4\,meV). The long recovery time ($>$80 ps) observed in $\Delta$R($\tau$)/R is reminiscent of heat-diffusion and can be associated with the long-lived pump-induced lattice distortions. These distortions would give rise to persistent band-structure modifications such as those observed with our TR-ARPES measurements. \\
We now compare TR-ARPES and TR-reflectivity data. One should note that the TSS dispersion modification takes place within our TR-ARPES system's temporal resolution (250\,fs), comparable to the period of the A$^{2}_{1g}$ optical phonon and the rise time of $\Delta$R($\tau$)/R (Fig.\,\ref{Fig2}(b-d)). However, while the A$^{2}_{1g}$ phonon decays within 1\,ps, the TSS dispersion is changed for substantially longer. The A$^{1}_{1g}$ phonon lasts instead for several picoseconds with a damping time comparable to that of the Fermi velocity and the Fermi momentum (Fig.\,\ref{Fig2}(c)), playing perhaps a more substantial role in the modification of the observed dispersion. Furthermore, additional modes not observed via reflectivity may also play a secondary role in stabilizing the modified dispersion over the long time scales measured in our experiment \cite{SHG_phonon_TI}. We can conclude with confidence that the modifications to the TSS dispersion can be attributed to a complex interplay of pump-induced phonons.

\section{Discussion\label{Discussion}}
With this modification to both the lattice and electronic structure thus confirmed via photoexcitation of several phonon modes, we can now address the resulting modification to the ARPES matrix elements which ensue. To do so, we focus our attention on the temporal evolution of several well-defined regions of energy and momentum along the different branches of the TSS. While the ultrafast scattering processes involving the TSS of several TIs have been reported \cite{ShenBiSe2012,SterziTI,BuginiBiSe,SobotaJElectr2014,PerfettiBiTeNatComm,PerfettiBiTe,TIgiganticlifetime_Hasan,HoferTHZpumpTI,RaderTR_SPIN_ARPES,Lanzara_SPIN_ARPES,WarpingBiTeRader2014}, the effect of the transient deformation potential discussed above on the conclusions of those studies has not been addressed. In Fig.\,\ref{Fig3}(a) we plot the TSS at +0.6 ps pump-probe delay for two different linear probe polarizations (s and p). The area of the integration region was chosen to be comparable in energy to our system resolution (20\,meV), and large enough in momentum so as to ensure that no states move in or out of the window with the change in dispersion. They are indicated by the colored boxes in Fig.\,\ref{Fig3}(a). It is important to note that as the Dirac point does not shift with the excitation (see Fig.\,\ref{Fig2}), the energy window is fixed with respect to the TSS for all time delays. The temporal evolution of the integrated intensity within these boxes, $\Delta I(\tau)$=$\int_{\omega,\textbf{k}} I_{\text{PES}}(\omega,\textbf{k},\tau) d \omega d \textbf{k}$, is then plotted in Fig.\,\ref{Fig3}(b). Comparing this evolution for different energy windows on both the left (L) and right (R)  branches of the Dirac cone with s- and p-polarized probe light, we find that, remarkably, $\Delta I(\tau)$ depends on the choice of probe polarization. We reiterate here that throughout, the pump polarization is fixed, and so this cannot be interpreted as the result of different excitations associated with a choice of pump polarization. As we probe only a single TSS within a given integration window, the observed differences in relaxation rates cannot be attributed to either $f(\omega,\textbf{k},\tau)$ or $A(\omega,\textbf{k},\tau)$, as these have no connection within this context to the probe pulse polarization. Rather it would seem that the matrix element factor exhibits distinct temporal evolution which depends on the choice of probe polarization. 
This finding is consistent with the model discussed in Fig.\,\ref{Fig1}(a). In particular, a simple exponential fit (for $\tau >$3\,ps) of the $\Delta I(\tau)$ curves at +25\,meV binding energy, left branch, results in the extraction of conflicting relaxation times of 4.6$\pm$0.3\,ps and 3.2$\pm$0.2\,ps for p- and s-polarized probe, respectively.\\
To isolate the dynamics of the matrix element factor, we introduce the quantity
\begin{equation} \label{EQ:2}
   \text{Z}(\tau)=(I_p/I_s)(\tau)=|M_p(\tau)|^2 / |M_s(\tau)|^2,
\end{equation}
defined as the ratio between photoemission intensities with p- and s-polarized light. 
The form of $Z(\tau)$ has been chosen to eliminate contributions from both $A(\omega,\textbf{k},\tau)$ and $f(\omega,\textbf{k},\tau)$, retaining only the relative matrix element dynamics. In the absence of temporal evolution for $M$, or for equivalent time dependence in both polarization channels, $Z(\tau)$ would be constant.
In Fig. 3(c), we plot $Z(\tau)$ as a function of the pump-probe delay for three different binding energies: two above the equilibrium chemical potential (+50 and +25 meV) and one below (-30 meV).
As exemplified by the lower panel, the matrix elements undergo an ultrafast response and subsequent relaxation following interaction with the pump. For all the three binding energies, we observe a transient evolution of $Z(\tau)$, unambiguously related to a dynamical matrix element ratio. \\
We also observe a strong dependence of $Z(\tau)$ on both momentum and energy, with the distinction between $s$ and $p$ polarized light seen primarily along the left branch. Ultimately, the microscopic origin of the dynamical matrix elements in BSTS is beyond the scope of this current work, as our primary objective is simply to demonstrate the important consequences of $Z(\tau)$ when characterizing the ultrafast response of the spectral and distribution functions. The particular form of $Z(\tau)$ here could arise due to pump-induced modifications to the initial state wavefunction, warping of the Dirac cone, or even the nature of the photoemission final states.
However, we note that the commonly reported photo-induced A$_{1g}$ optical phonons in TIs (see Fig.\,\ref{Fig2}(d-e) and Refs.\,\cite{ShenBiSe2014,BoschiniSciRep,Richter1977}) are out-of-plane modes, \emph{i.e.} along the c-axis. These out-of-plane phonons may induce a transient modification of the relative distance between atomic-layers, modulating the quantum interference effects. Such changes may result in an energy- and momentum-dependent modification of the photoemission intensity \cite{DamascelliBiSe2013,DamascelliBiSe2014,Ast-graphene-interference,Chiang-graphene}, similar to what has been reported here (see Fig\,\ref{Fig3}).  
The multitude of plausible contributions to $Z(\tau)$ emphasizes the theoretical challenge presented by dynamical matrix elements in TR-ARPES experiments; a challenge which needs be addressed in order to ensure successful application of this technique in the quantitative study of topological insulators and other materials. 

\section{Conclusion\label{Conclusion}}
We have reported a substantial photoinduced modification of the electronic structure of the topological insulator BSTS. This response is manifest in corrections to both the electronic dispersion and eigenstates. We discussed the scenario where both $A(\omega,\textbf{k},\tau)$ and $|M_{f,i}^{\textbf{k}}(\tau)|^2$ display dynamical behaviour, and the implications of this for the interpretation of TR-ARPES experiments.  We showed here that a probe-polarization study can be used to establish time-dependence within the different $|M_{f,i}^{\textbf{k}}(\tau)|^2$ channels.  By isolating the dynamics of the dipole matrix elements, one may hope to achieve a comprehensive understanding of the non-equilibrium properties of complex solid state systems.

\section{Acknowledgments}
We gratefully thank H.-H. Kung for fruitful discussions.
This research was undertaken thanks in part to funding from the Max Planck-UBC-UTokyo Centre for Quantum Materials and the Canada First Research Excellence Fund, Quantum Materials and Future Technologies Program. The work at UBC was supported by the Gordon and Betty Moore Foundation's  EPiQS Initiative, Grant GBMF4779, the Killam, Alfred P. Sloan, and Natural Sciences and Engineering Research Council of Canada's (NSERC's) Steacie Memorial Fellowships (A.D.), the Alexander von Humboldt Fellowship (A.D.), the Canada Research Chairs Program (A.D.), NSERC, Canada Foundation for Innovation (CFI), CIFAR Quantum Materials, and CIFAR Global Scholars (E.H.d.S.N.). E.R. acknowledges support from the Swiss National Science Foundation (SNSF) grant no. P300P2\_164649.
C.G. acknowledges financial support from MIUR through the PRIN 2015 Programme (Prot. 2015C5SEJJ001) and from Universit\`{a} Cattolica del Sacro Cuore through D.1, D.2.2 and D.3.1 grants.
This project has received funding from the European Union's Horizon 2020 research and innovation programme under grant agreement 785219 GrapheneCore2.
S.K.K. acknowledges the Laboratory Directed Research and Development (LDRD)--XWVM-LABR0000 and US Department of Energy Office of Science, BESMSE Science of 100 Tesla programs.\\

\providecommand{\noopsort}[1]{}\providecommand{\singleletter}[1]{#1}


\begin{thebibliography}{40}%
\makeatletter
\providecommand \@ifxundefined [1]{%
 \@ifx{#1\undefined}
}%
\providecommand \@ifnum [1]{%
 \ifnum #1\expandafter \@firstoftwo
 \else \expandafter \@secondoftwo
 \fi
}%
\providecommand \@ifx [1]{%
 \ifx #1\expandafter \@firstoftwo
 \else \expandafter \@secondoftwo
 \fi
}%
\providecommand \natexlab [1]{#1}%
\providecommand \enquote  [1]{``#1''}%
\providecommand \bibnamefont  [1]{#1}%
\providecommand \bibfnamefont [1]{#1}%
\providecommand \citenamefont [1]{#1}%
\providecommand \href@noop [0]{\@secondoftwo}%
\providecommand \href [0]{\begingroup \@sanitize@url \@href}%
\providecommand \@href[1]{\@@startlink{#1}\@@href}%
\providecommand \@@href[1]{\endgroup#1\@@endlink}%
\providecommand \@sanitize@url [0]{\catcode `\\12\catcode `\$12\catcode
  `\&12\catcode `\#12\catcode `\^12\catcode `\_12\catcode `\%12\relax}%
\providecommand \@@startlink[1]{}%
\providecommand \@@endlink[0]{}%
\providecommand \url  [0]{\begingroup\@sanitize@url \@url }%
\providecommand \@url [1]{\endgroup\@href {#1}{\urlprefix }}%
\providecommand \urlprefix  [0]{URL }%
\providecommand \Eprint [0]{\href }%
\providecommand \doibase [0]{http://dx.doi.org/}%
\providecommand \selectlanguage [0]{\@gobble}%
\providecommand \bibinfo  [0]{\@secondoftwo}%
\providecommand \bibfield  [0]{\@secondoftwo}%
\providecommand \translation [1]{[#1]}%
\providecommand \BibitemOpen [0]{}%
\providecommand \bibitemStop [0]{}%
\providecommand \bibitemNoStop [0]{.\EOS\space}%
\providecommand \EOS [0]{\spacefactor3000\relax}%
\providecommand \BibitemShut  [1]{\csname bibitem#1\endcsname}%
\let\auto@bib@innerbib\@empty
\bibitem [{\citenamefont {Giannetti}\ \emph {et~al.}(2016)\citenamefont
  {Giannetti}, \citenamefont {Capone}, \citenamefont {Fausti}, \citenamefont
  {Fabrizio}, \citenamefont {Parmigiani},\ and\ \citenamefont
  {Mihailovic}}]{ReviewGiannetti}%
  \BibitemOpen
  \bibfield  {author} {\bibinfo {author} {\bibfnamefont {C.}~\bibnamefont
  {Giannetti}}, \bibinfo {author} {\bibfnamefont {M.}~\bibnamefont {Capone}},
  \bibinfo {author} {\bibfnamefont {D.}~\bibnamefont {Fausti}}, \bibinfo
  {author} {\bibfnamefont {M.}~\bibnamefont {Fabrizio}}, \bibinfo {author}
  {\bibfnamefont {F.}~\bibnamefont {Parmigiani}}, \ and\ \bibinfo {author}
  {\bibfnamefont {D.}~\bibnamefont {Mihailovic}},\ }\href {\doibase
  10.1080/00018732.2016.1194044} {\bibfield  {journal} {\bibinfo  {journal}
  {Advances in Physics}\ }\textbf {\bibinfo {volume} {65}},\ \bibinfo {pages}
  {58} (\bibinfo {year} {2016})}\BibitemShut {NoStop}%
\bibitem [{\citenamefont {Boschini}\ \emph {et~al.}(2018)\citenamefont
  {Boschini}, \citenamefont {da~Silva~Neto}, \citenamefont {Razzoli},
  \citenamefont {Zonno}, \citenamefont {Peli}, \citenamefont {Day},
  \citenamefont {Michiardi}, \citenamefont {Schneider}, \citenamefont
  {Zwartsenberg}, \citenamefont {Nigge}, \citenamefont {Zhong}, \citenamefont
  {Schneeloch}, \citenamefont {Gu}, \citenamefont {Zhdanovich}, \citenamefont
  {Mills}, \citenamefont {Levy}, \citenamefont {Jones},\ and\ \citenamefont
  {Damascelli}}]{Boschini2018}%
  \BibitemOpen
  \bibfield  {author} {\bibinfo {author} {\bibfnamefont {F.}~\bibnamefont
  {Boschini}}, \bibinfo {author} {\bibfnamefont {E.~H.}\ \bibnamefont
  {da~Silva~Neto}}, \bibinfo {author} {\bibfnamefont {E.}~\bibnamefont
  {Razzoli}}, \bibinfo {author} {\bibfnamefont {M.}~\bibnamefont {Zonno}},
  \bibinfo {author} {\bibfnamefont {S.}~\bibnamefont {Peli}}, \bibinfo {author}
  {\bibfnamefont {R.~P.}\ \bibnamefont {Day}}, \bibinfo {author} {\bibfnamefont
  {M.}~\bibnamefont {Michiardi}}, \bibinfo {author} {\bibfnamefont
  {M.}~\bibnamefont {Schneider}}, \bibinfo {author} {\bibfnamefont
  {B.}~\bibnamefont {Zwartsenberg}}, \bibinfo {author} {\bibfnamefont
  {P.}~\bibnamefont {Nigge}}, \bibinfo {author} {\bibfnamefont {R.~D.}\
  \bibnamefont {Zhong}}, \bibinfo {author} {\bibfnamefont {J.}~\bibnamefont
  {Schneeloch}}, \bibinfo {author} {\bibfnamefont {G.~D.}\ \bibnamefont {Gu}},
  \bibinfo {author} {\bibfnamefont {S.}~\bibnamefont {Zhdanovich}}, \bibinfo
  {author} {\bibfnamefont {A.~K.}\ \bibnamefont {Mills}}, \bibinfo {author}
  {\bibfnamefont {G.}~\bibnamefont {Levy}}, \bibinfo {author} {\bibfnamefont
  {C.}~\bibnamefont {Jones}, \bibfnamefont {D.~J.~andGiannetti}}, \ and\
  \bibinfo {author} {\bibfnamefont {A.}~\bibnamefont {Damascelli}},\ }\href
  {\doibase 10.1038/s41563-018-0045-1} {\bibfield  {journal} {\bibinfo
  {journal} {Nature Materials}\ }\textbf {\bibinfo {volume} {17}},\ \bibinfo
  {pages} {416} (\bibinfo {year} {2018})}\BibitemShut {NoStop}%
\bibitem [{\citenamefont {Smallwood}\ \emph {et~al.}(2012)\citenamefont
  {Smallwood}, \citenamefont {Hinton}, \citenamefont {Jozwiak}, \citenamefont
  {Zhang}, \citenamefont {Koralek}, \citenamefont {Eisaki}, \citenamefont
  {Lee}, \citenamefont {Orenstein},\ and\ \citenamefont
  {Lanzara}}]{ScienceLanzara}%
  \BibitemOpen
  \bibfield  {author} {\bibinfo {author} {\bibfnamefont {C.~L.}\ \bibnamefont
  {Smallwood}}, \bibinfo {author} {\bibfnamefont {J.~P.}\ \bibnamefont
  {Hinton}}, \bibinfo {author} {\bibfnamefont {C.}~\bibnamefont {Jozwiak}},
  \bibinfo {author} {\bibfnamefont {W.}~\bibnamefont {Zhang}}, \bibinfo
  {author} {\bibfnamefont {J.~D.}\ \bibnamefont {Koralek}}, \bibinfo {author}
  {\bibfnamefont {H.}~\bibnamefont {Eisaki}}, \bibinfo {author} {\bibfnamefont
  {D.-H.}\ \bibnamefont {Lee}}, \bibinfo {author} {\bibfnamefont
  {J.}~\bibnamefont {Orenstein}}, \ and\ \bibinfo {author} {\bibfnamefont
  {A.}~\bibnamefont {Lanzara}},\ }\href {\doibase 10.1126/science.1217423}
  {\bibfield  {journal} {\bibinfo  {journal} {Science}\ }\textbf {\bibinfo
  {volume} {336}},\ \bibinfo {pages} {1137} (\bibinfo {year}
  {2012})}\BibitemShut {NoStop}%
\bibitem [{\citenamefont {Rettig}\ \emph {et~al.}(2013)\citenamefont {Rettig},
  \citenamefont {Cortés}, \citenamefont {Jeevan}, \citenamefont {Gegenwart},
  \citenamefont {Wolf}, \citenamefont {Fink},\ and\ \citenamefont
  {Bovensiepen}}]{BovensiepenFebased}%
  \BibitemOpen
  \bibfield  {author} {\bibinfo {author} {\bibfnamefont {L.}~\bibnamefont
  {Rettig}}, \bibinfo {author} {\bibfnamefont {R.}~\bibnamefont {Cortés}},
  \bibinfo {author} {\bibfnamefont {H.~S.}\ \bibnamefont {Jeevan}}, \bibinfo
  {author} {\bibfnamefont {P.}~\bibnamefont {Gegenwart}}, \bibinfo {author}
  {\bibfnamefont {T.}~\bibnamefont {Wolf}}, \bibinfo {author} {\bibfnamefont
  {J.}~\bibnamefont {Fink}}, \ and\ \bibinfo {author} {\bibfnamefont
  {U.}~\bibnamefont {Bovensiepen}},\ }\href
  {http://stacks.iop.org/1367-2630/15/i=8/a=083023} {\bibfield  {journal}
  {\bibinfo  {journal} {New Journal of Physics}\ }\textbf {\bibinfo {volume}
  {15}},\ \bibinfo {pages} {083023} (\bibinfo {year} {2013})}\BibitemShut
  {NoStop}%
\bibitem [{\citenamefont {Rohwer}\ \emph {et~al.}(2011)\citenamefont {Rohwer},
  \citenamefont {Hellmann}, \citenamefont {Wiesenmayer}, \citenamefont {Sohrt},
  \citenamefont {Stange}, \citenamefont {Slomski}, \citenamefont {Carr},
  \citenamefont {Liu}, \citenamefont {Avila}, \citenamefont {Kalläne},
  \citenamefont {Mathias}, \citenamefont {Kipp}, \citenamefont {Rossnagel},\
  and\ \citenamefont {Bauer}}]{CollapseCDWTiSe2}%
  \BibitemOpen
  \bibfield  {author} {\bibinfo {author} {\bibfnamefont {T.}~\bibnamefont
  {Rohwer}}, \bibinfo {author} {\bibfnamefont {S.}~\bibnamefont {Hellmann}},
  \bibinfo {author} {\bibfnamefont {M.}~\bibnamefont {Wiesenmayer}}, \bibinfo
  {author} {\bibfnamefont {C.}~\bibnamefont {Sohrt}}, \bibinfo {author}
  {\bibfnamefont {A.}~\bibnamefont {Stange}}, \bibinfo {author} {\bibfnamefont
  {B.}~\bibnamefont {Slomski}}, \bibinfo {author} {\bibfnamefont
  {A.}~\bibnamefont {Carr}}, \bibinfo {author} {\bibfnamefont {Y.}~\bibnamefont
  {Liu}}, \bibinfo {author} {\bibfnamefont {L.~M.}\ \bibnamefont {Avila}},
  \bibinfo {author} {\bibfnamefont {M.}~\bibnamefont {Kalläne}}, \bibinfo
  {author} {\bibfnamefont {S.}~\bibnamefont {Mathias}}, \bibinfo {author}
  {\bibfnamefont {L.}~\bibnamefont {Kipp}}, \bibinfo {author} {\bibfnamefont
  {K.}~\bibnamefont {Rossnagel}}, \ and\ \bibinfo {author} {\bibfnamefont
  {M.}~\bibnamefont {Bauer}},\ }\href {\doibase 10.1038/nature09829} {\bibfield
   {journal} {\bibinfo  {journal} {Nature}\ }\textbf {\bibinfo {volume}
  {471}},\ \bibinfo {pages} {490} (\bibinfo {year} {2011})}\BibitemShut
  {NoStop}%
\bibitem [{\citenamefont {Mor}\ \emph {et~al.}(2017)\citenamefont {Mor},
  \citenamefont {Herzog}, \citenamefont {Gole\ifmmode~\check{z}\else
  \v{z}\fi{}}, \citenamefont {Werner}, \citenamefont {Eckstein}, \citenamefont
  {Katayama}, \citenamefont {Nohara}, \citenamefont {Takagi}, \citenamefont
  {Mizokawa}, \citenamefont {Monney},\ and\ \citenamefont
  {St\"ahler}}]{StahlerExcitonic}%
  \BibitemOpen
  \bibfield  {author} {\bibinfo {author} {\bibfnamefont {S.}~\bibnamefont
  {Mor}}, \bibinfo {author} {\bibfnamefont {M.}~\bibnamefont {Herzog}},
  \bibinfo {author} {\bibfnamefont {D.}~\bibnamefont
  {Gole\ifmmode~\check{z}\else \v{z}\fi{}}}, \bibinfo {author} {\bibfnamefont
  {P.}~\bibnamefont {Werner}}, \bibinfo {author} {\bibfnamefont
  {M.}~\bibnamefont {Eckstein}}, \bibinfo {author} {\bibfnamefont
  {N.}~\bibnamefont {Katayama}}, \bibinfo {author} {\bibfnamefont
  {M.}~\bibnamefont {Nohara}}, \bibinfo {author} {\bibfnamefont
  {H.}~\bibnamefont {Takagi}}, \bibinfo {author} {\bibfnamefont
  {T.}~\bibnamefont {Mizokawa}}, \bibinfo {author} {\bibfnamefont
  {C.}~\bibnamefont {Monney}}, \ and\ \bibinfo {author} {\bibfnamefont
  {J.}~\bibnamefont {St\"ahler}},\ }\href {\doibase
  10.1103/PhysRevLett.119.086401} {\bibfield  {journal} {\bibinfo  {journal}
  {Phys. Rev. Lett.}\ }\textbf {\bibinfo {volume} {119}},\ \bibinfo {pages}
  {086401} (\bibinfo {year} {2017})}\BibitemShut {NoStop}%
\bibitem [{\citenamefont {Wang}\ \emph {et~al.}(2013)\citenamefont {Wang},
  \citenamefont {Steinberg}, \citenamefont {Jarillo-Herrero},\ and\
  \citenamefont {Gedik}}]{GedikFloquetScience}%
  \BibitemOpen
  \bibfield  {author} {\bibinfo {author} {\bibfnamefont {Y.~H.}\ \bibnamefont
  {Wang}}, \bibinfo {author} {\bibfnamefont {H.}~\bibnamefont {Steinberg}},
  \bibinfo {author} {\bibfnamefont {P.}~\bibnamefont {Jarillo-Herrero}}, \ and\
  \bibinfo {author} {\bibfnamefont {N.}~\bibnamefont {Gedik}},\ }\href
  {\doibase 10.1126/science.1239834} {\bibfield  {journal} {\bibinfo  {journal}
  {Science}\ }\textbf {\bibinfo {volume} {342}},\ \bibinfo {pages} {453}
  (\bibinfo {year} {2013})}\BibitemShut {NoStop}%
\bibitem [{\citenamefont {Mahmood}\ \emph {et~al.}(2016)\citenamefont
  {Mahmood}, \citenamefont {Chan}, \citenamefont {Alpichshev}, \citenamefont
  {Gardner}, \citenamefont {Lee}, \citenamefont {Lee},\ and\ \citenamefont
  {Gedik}}]{GedikFloquetNatPhys}%
  \BibitemOpen
  \bibfield  {author} {\bibinfo {author} {\bibfnamefont {F.}~\bibnamefont
  {Mahmood}}, \bibinfo {author} {\bibfnamefont {C.-K.}\ \bibnamefont {Chan}},
  \bibinfo {author} {\bibfnamefont {Z.}~\bibnamefont {Alpichshev}}, \bibinfo
  {author} {\bibfnamefont {D.}~\bibnamefont {Gardner}}, \bibinfo {author}
  {\bibfnamefont {Y.}~\bibnamefont {Lee}}, \bibinfo {author} {\bibfnamefont
  {P.~A.}\ \bibnamefont {Lee}}, \ and\ \bibinfo {author} {\bibfnamefont
  {N.}~\bibnamefont {Gedik}},\ }\href {\doibase 10.1038/nphys3609} {\bibfield
  {journal} {\bibinfo  {journal} {Nature Physics}\ }\textbf {\bibinfo {volume}
  {12}},\ \bibinfo {pages} {306} (\bibinfo {year} {2016})}\BibitemShut
  {NoStop}%
\bibitem [{\citenamefont {Sterzi}\ \emph {et~al.}(2016)\citenamefont {Sterzi},
  \citenamefont {Crepaldi}, \citenamefont {Cilento}, \citenamefont {Manzoni},
  \citenamefont {Frantzeskakis}, \citenamefont {Zacchigna}, \citenamefont {van
  Heumen}, \citenamefont {Huang}, \citenamefont {Golden},\ and\ \citenamefont
  {Parmigiani}}]{SterziSmB6}%
  \BibitemOpen
  \bibfield  {author} {\bibinfo {author} {\bibfnamefont {A.}~\bibnamefont
  {Sterzi}}, \bibinfo {author} {\bibfnamefont {A.}~\bibnamefont {Crepaldi}},
  \bibinfo {author} {\bibfnamefont {F.}~\bibnamefont {Cilento}}, \bibinfo
  {author} {\bibfnamefont {G.}~\bibnamefont {Manzoni}}, \bibinfo {author}
  {\bibfnamefont {E.}~\bibnamefont {Frantzeskakis}}, \bibinfo {author}
  {\bibfnamefont {M.}~\bibnamefont {Zacchigna}}, \bibinfo {author}
  {\bibfnamefont {E.}~\bibnamefont {van Heumen}}, \bibinfo {author}
  {\bibfnamefont {Y.~K.}\ \bibnamefont {Huang}}, \bibinfo {author}
  {\bibfnamefont {M.~S.}\ \bibnamefont {Golden}}, \ and\ \bibinfo {author}
  {\bibfnamefont {F.}~\bibnamefont {Parmigiani}},\ }\href {\doibase
  10.1103/PhysRevB.94.081111} {\bibfield  {journal} {\bibinfo  {journal} {Phys.
  Rev. B}\ }\textbf {\bibinfo {volume} {94}},\ \bibinfo {pages} {081111}
  (\bibinfo {year} {2016})}\BibitemShut {NoStop}%
\bibitem [{\citenamefont {Sterzi}\ \emph {et~al.}(2017)\citenamefont {Sterzi},
  \citenamefont {Manzoni}, \citenamefont {Sbuelz}, \citenamefont {Cilento},
  \citenamefont {Zacchigna}, \citenamefont {Bugnon}, \citenamefont {Magrez},
  \citenamefont {Berger}, \citenamefont {Crepaldi},\ and\ \citenamefont
  {Parmigiani}}]{SterziTI}%
  \BibitemOpen
  \bibfield  {author} {\bibinfo {author} {\bibfnamefont {A.}~\bibnamefont
  {Sterzi}}, \bibinfo {author} {\bibfnamefont {G.}~\bibnamefont {Manzoni}},
  \bibinfo {author} {\bibfnamefont {L.}~\bibnamefont {Sbuelz}}, \bibinfo
  {author} {\bibfnamefont {F.}~\bibnamefont {Cilento}}, \bibinfo {author}
  {\bibfnamefont {M.}~\bibnamefont {Zacchigna}}, \bibinfo {author}
  {\bibfnamefont {P.}~\bibnamefont {Bugnon}}, \bibinfo {author} {\bibfnamefont
  {A.}~\bibnamefont {Magrez}}, \bibinfo {author} {\bibfnamefont
  {H.}~\bibnamefont {Berger}}, \bibinfo {author} {\bibfnamefont
  {A.}~\bibnamefont {Crepaldi}}, \ and\ \bibinfo {author} {\bibfnamefont
  {F.}~\bibnamefont {Parmigiani}},\ }\href {\doibase
  10.1103/PhysRevB.95.115431} {\bibfield  {journal} {\bibinfo  {journal} {Phys.
  Rev. B}\ }\textbf {\bibinfo {volume} {95}},\ \bibinfo {pages} {115431}
  (\bibinfo {year} {2017})}\BibitemShut {NoStop}%
\bibitem [{\citenamefont {Perfetti}\ \emph {et~al.}(2007)\citenamefont
  {Perfetti}, \citenamefont {Loukakos}, \citenamefont {Lisowski}, \citenamefont
  {Bovensiepen}, \citenamefont {Eisaki},\ and\ \citenamefont
  {Wolf}}]{PRL2007PerfettiBi2212}%
  \BibitemOpen
  \bibfield  {author} {\bibinfo {author} {\bibfnamefont {L.}~\bibnamefont
  {Perfetti}}, \bibinfo {author} {\bibfnamefont {P.~A.}\ \bibnamefont
  {Loukakos}}, \bibinfo {author} {\bibfnamefont {M.}~\bibnamefont {Lisowski}},
  \bibinfo {author} {\bibfnamefont {U.}~\bibnamefont {Bovensiepen}}, \bibinfo
  {author} {\bibfnamefont {H.}~\bibnamefont {Eisaki}}, \ and\ \bibinfo {author}
  {\bibfnamefont {M.}~\bibnamefont {Wolf}},\ }\href {\doibase
  10.1103/PhysRevLett.99.197001} {\bibfield  {journal} {\bibinfo  {journal}
  {Phys. Rev. Lett.}\ }\textbf {\bibinfo {volume} {99}},\ \bibinfo {pages}
  {197001} (\bibinfo {year} {2007})}\BibitemShut {NoStop}%
\bibitem [{\citenamefont {Crepaldi}\ \emph {et~al.}(2017)\citenamefont
  {Crepaldi}, \citenamefont {Aut\`es}, \citenamefont {Gatti}, \citenamefont
  {Roth}, \citenamefont {Sterzi}, \citenamefont {Manzoni}, \citenamefont
  {Zacchigna}, \citenamefont {Cacho}, \citenamefont {Chapman}, \citenamefont
  {Springate}, \citenamefont {Seddon}, \citenamefont {Bugnon}, \citenamefont
  {Magrez}, \citenamefont {Berger}, \citenamefont {Vobornik}, \citenamefont
  {Kall\"ane}, \citenamefont {Quer}, \citenamefont {Rossnagel}, \citenamefont
  {Parmigiani}, \citenamefont {Yazyev},\ and\ \citenamefont
  {Grioni}}]{CrepaldiWeyl}%
  \BibitemOpen
  \bibfield  {author} {\bibinfo {author} {\bibfnamefont {A.}~\bibnamefont
  {Crepaldi}}, \bibinfo {author} {\bibfnamefont {G.}~\bibnamefont {Aut\`es}},
  \bibinfo {author} {\bibfnamefont {G.}~\bibnamefont {Gatti}}, \bibinfo
  {author} {\bibfnamefont {S.}~\bibnamefont {Roth}}, \bibinfo {author}
  {\bibfnamefont {A.}~\bibnamefont {Sterzi}}, \bibinfo {author} {\bibfnamefont
  {G.}~\bibnamefont {Manzoni}}, \bibinfo {author} {\bibfnamefont
  {M.}~\bibnamefont {Zacchigna}}, \bibinfo {author} {\bibfnamefont
  {C.}~\bibnamefont {Cacho}}, \bibinfo {author} {\bibfnamefont {R.~T.}\
  \bibnamefont {Chapman}}, \bibinfo {author} {\bibfnamefont {E.}~\bibnamefont
  {Springate}}, \bibinfo {author} {\bibfnamefont {E.~A.}\ \bibnamefont
  {Seddon}}, \bibinfo {author} {\bibfnamefont {P.}~\bibnamefont {Bugnon}},
  \bibinfo {author} {\bibfnamefont {A.}~\bibnamefont {Magrez}}, \bibinfo
  {author} {\bibfnamefont {H.}~\bibnamefont {Berger}}, \bibinfo {author}
  {\bibfnamefont {I.}~\bibnamefont {Vobornik}}, \bibinfo {author}
  {\bibfnamefont {M.}~\bibnamefont {Kall\"ane}}, \bibinfo {author}
  {\bibfnamefont {A.}~\bibnamefont {Quer}}, \bibinfo {author} {\bibfnamefont
  {K.}~\bibnamefont {Rossnagel}}, \bibinfo {author} {\bibfnamefont
  {F.}~\bibnamefont {Parmigiani}}, \bibinfo {author} {\bibfnamefont {O.~V.}\
  \bibnamefont {Yazyev}}, \ and\ \bibinfo {author} {\bibfnamefont
  {M.}~\bibnamefont {Grioni}},\ }\href {\doibase 10.1103/PhysRevB.96.241408}
  {\bibfield  {journal} {\bibinfo  {journal} {Phys. Rev. B}\ }\textbf {\bibinfo
  {volume} {96}},\ \bibinfo {pages} {241408} (\bibinfo {year}
  {2017})}\BibitemShut {NoStop}%
\bibitem [{\citenamefont {Hajlaoui}\ \emph {et~al.}(2012)\citenamefont
  {Hajlaoui}, \citenamefont {Papalazarou}, \citenamefont {Mauchain},
  \citenamefont {Lantz}, \citenamefont {Moisan}, \citenamefont {Boschetto},
  \citenamefont {Jiang}, \citenamefont {Miotkowski}, \citenamefont {Chen},
  \citenamefont {Taleb-Ibrahimi}, \citenamefont {Perfetti},\ and\ \citenamefont
  {Marsi}}]{PerfettiBiTe}%
  \BibitemOpen
  \bibfield  {author} {\bibinfo {author} {\bibfnamefont {M.}~\bibnamefont
  {Hajlaoui}}, \bibinfo {author} {\bibfnamefont {E.}~\bibnamefont
  {Papalazarou}}, \bibinfo {author} {\bibfnamefont {J.}~\bibnamefont
  {Mauchain}}, \bibinfo {author} {\bibfnamefont {G.}~\bibnamefont {Lantz}},
  \bibinfo {author} {\bibfnamefont {N.}~\bibnamefont {Moisan}}, \bibinfo
  {author} {\bibfnamefont {D.}~\bibnamefont {Boschetto}}, \bibinfo {author}
  {\bibfnamefont {Z.}~\bibnamefont {Jiang}}, \bibinfo {author} {\bibfnamefont
  {I.}~\bibnamefont {Miotkowski}}, \bibinfo {author} {\bibfnamefont {Y.~P.}\
  \bibnamefont {Chen}}, \bibinfo {author} {\bibfnamefont {A.}~\bibnamefont
  {Taleb-Ibrahimi}}, \bibinfo {author} {\bibfnamefont {L.}~\bibnamefont
  {Perfetti}}, \ and\ \bibinfo {author} {\bibfnamefont {M.}~\bibnamefont
  {Marsi}},\ }\href {\doibase 10.1021/nl301035x} {\bibfield  {journal}
  {\bibinfo  {journal} {Nano Letters}\ }\textbf {\bibinfo {volume} {12}},\
  \bibinfo {pages} {3532} (\bibinfo {year} {2012})}\BibitemShut {NoStop}%
\bibitem [{\citenamefont {Sobota}\ \emph {et~al.}(2012)\citenamefont {Sobota},
  \citenamefont {Yang}, \citenamefont {Analytis}, \citenamefont {Chen},
  \citenamefont {Fisher}, \citenamefont {Kirchmann},\ and\ \citenamefont
  {Shen}}]{ShenBiSe2012}%
  \BibitemOpen
  \bibfield  {author} {\bibinfo {author} {\bibfnamefont {J.~A.}\ \bibnamefont
  {Sobota}}, \bibinfo {author} {\bibfnamefont {S.}~\bibnamefont {Yang}},
  \bibinfo {author} {\bibfnamefont {J.~G.}\ \bibnamefont {Analytis}}, \bibinfo
  {author} {\bibfnamefont {Y.~L.}\ \bibnamefont {Chen}}, \bibinfo {author}
  {\bibfnamefont {I.~R.}\ \bibnamefont {Fisher}}, \bibinfo {author}
  {\bibfnamefont {P.~S.}\ \bibnamefont {Kirchmann}}, \ and\ \bibinfo {author}
  {\bibfnamefont {Z.-X.}\ \bibnamefont {Shen}},\ }\href {\doibase
  10.1103/PhysRevLett.108.117403} {\bibfield  {journal} {\bibinfo  {journal}
  {Phys. Rev. Lett.}\ }\textbf {\bibinfo {volume} {108}},\ \bibinfo {pages}
  {117403} (\bibinfo {year} {2012})}\BibitemShut {NoStop}%
\bibitem [{\citenamefont {Bugini}\ \emph {et~al.}(2017)\citenamefont {Bugini},
  \citenamefont {Boschini}, \citenamefont {Hedayat}, \citenamefont {Yi},
  \citenamefont {Chen}, \citenamefont {Zhou}, \citenamefont {Manzoni},
  \citenamefont {Dallera}, \citenamefont {Cerullo},\ and\ \citenamefont
  {Carpene}}]{BuginiBiSe}%
  \BibitemOpen
  \bibfield  {author} {\bibinfo {author} {\bibfnamefont {D.}~\bibnamefont
  {Bugini}}, \bibinfo {author} {\bibfnamefont {F.}~\bibnamefont {Boschini}},
  \bibinfo {author} {\bibfnamefont {H.}~\bibnamefont {Hedayat}}, \bibinfo
  {author} {\bibfnamefont {H.}~\bibnamefont {Yi}}, \bibinfo {author}
  {\bibfnamefont {C.}~\bibnamefont {Chen}}, \bibinfo {author} {\bibfnamefont
  {X.}~\bibnamefont {Zhou}}, \bibinfo {author} {\bibfnamefont {C.}~\bibnamefont
  {Manzoni}}, \bibinfo {author} {\bibfnamefont {C.}~\bibnamefont {Dallera}},
  \bibinfo {author} {\bibfnamefont {G.}~\bibnamefont {Cerullo}}, \ and\
  \bibinfo {author} {\bibfnamefont {E.}~\bibnamefont {Carpene}},\ }\href
  {http://stacks.iop.org/0953-8984/29/i=30/a=30LT01} {\bibfield  {journal}
  {\bibinfo  {journal} {Journal of Physics: Condensed Matter}\ }\textbf
  {\bibinfo {volume} {29}},\ \bibinfo {pages} {30LT01} (\bibinfo {year}
  {2017})}\BibitemShut {NoStop}%
\bibitem [{\citenamefont {Damascelli}(2004)}]{ReviewDamascelli}%
  \BibitemOpen
  \bibfield  {author} {\bibinfo {author} {\bibfnamefont {A.}~\bibnamefont
  {Damascelli}},\ }\href@noop {} {\bibfield  {journal} {\bibinfo  {journal}
  {Physica Scripta}\ }\textbf {\bibinfo {volume} {2004}},\ \bibinfo {pages}
  {61} (\bibinfo {year} {2004})}\BibitemShut {NoStop}%
\bibitem [{\citenamefont {Randi}\ \emph {et~al.}(2017)\citenamefont {Randi},
  \citenamefont {Fausti},\ and\ \citenamefont
  {Eckstein}}]{FaustiBeyondEnergyRes}%
  \BibitemOpen
  \bibfield  {author} {\bibinfo {author} {\bibfnamefont {F.}~\bibnamefont
  {Randi}}, \bibinfo {author} {\bibfnamefont {D.}~\bibnamefont {Fausti}}, \
  and\ \bibinfo {author} {\bibfnamefont {M.}~\bibnamefont {Eckstein}},\ }\href
  {\doibase 10.1103/PhysRevB.95.115132} {\bibfield  {journal} {\bibinfo
  {journal} {Phys. Rev. B}\ }\textbf {\bibinfo {volume} {95}},\ \bibinfo
  {pages} {115132} (\bibinfo {year} {2017})}\BibitemShut {NoStop}%
\bibitem [{\citenamefont {Sobota}\ \emph
  {et~al.}(2014{\natexlab{a}})\citenamefont {Sobota}, \citenamefont {Yang},
  \citenamefont {Leuenberger}, \citenamefont {Kemper}, \citenamefont
  {Analytis}, \citenamefont {Fisher}, \citenamefont {Kirchmann}, \citenamefont
  {Devereaux},\ and\ \citenamefont {Shen}}]{ShenBiSe2014}%
  \BibitemOpen
  \bibfield  {author} {\bibinfo {author} {\bibfnamefont {J.~A.}\ \bibnamefont
  {Sobota}}, \bibinfo {author} {\bibfnamefont {S.-L.}\ \bibnamefont {Yang}},
  \bibinfo {author} {\bibfnamefont {D.}~\bibnamefont {Leuenberger}}, \bibinfo
  {author} {\bibfnamefont {A.~F.}\ \bibnamefont {Kemper}}, \bibinfo {author}
  {\bibfnamefont {J.~G.}\ \bibnamefont {Analytis}}, \bibinfo {author}
  {\bibfnamefont {I.~R.}\ \bibnamefont {Fisher}}, \bibinfo {author}
  {\bibfnamefont {P.~S.}\ \bibnamefont {Kirchmann}}, \bibinfo {author}
  {\bibfnamefont {T.~P.}\ \bibnamefont {Devereaux}}, \ and\ \bibinfo {author}
  {\bibfnamefont {Z.-X.}\ \bibnamefont {Shen}},\ }\href {\doibase
  10.1103/PhysRevLett.113.157401} {\bibfield  {journal} {\bibinfo  {journal}
  {Phys. Rev. Lett.}\ }\textbf {\bibinfo {volume} {113}},\ \bibinfo {pages}
  {157401} (\bibinfo {year} {2014}{\natexlab{a}})}\BibitemShut {NoStop}%
\bibitem [{\citenamefont {Gerber}\ \emph {et~al.}(2017)\citenamefont {Gerber},
  \citenamefont {Yang}, \citenamefont {Zhu}, \citenamefont {Soifer},
  \citenamefont {Sobota}, \citenamefont {Rebec}, \citenamefont {Lee},
  \citenamefont {Jia}, \citenamefont {Moritz}, \citenamefont {Jia},
  \citenamefont {Gauthier}, \citenamefont {Li}, \citenamefont {Leuenberger},
  \citenamefont {Zhang}, \citenamefont {Chaix}, \citenamefont {Li},
  \citenamefont {Jang}, \citenamefont {Lee}, \citenamefont {Yi}, \citenamefont
  {Dakovski}, \citenamefont {Song}, \citenamefont {Glownia}, \citenamefont
  {Nelson}, \citenamefont {Kim}, \citenamefont {Chuang}, \citenamefont
  {Hussain}, \citenamefont {Moore}, \citenamefont {Devereaux}, \citenamefont
  {Lee}, \citenamefont {Kirchmann},\ and\ \citenamefont
  {Shen}}]{ScienceShenTRxrd}%
  \BibitemOpen
  \bibfield  {author} {\bibinfo {author} {\bibfnamefont {S.}~\bibnamefont
  {Gerber}}, \bibinfo {author} {\bibfnamefont {S.-L.}\ \bibnamefont {Yang}},
  \bibinfo {author} {\bibfnamefont {D.}~\bibnamefont {Zhu}}, \bibinfo {author}
  {\bibfnamefont {H.}~\bibnamefont {Soifer}}, \bibinfo {author} {\bibfnamefont
  {J.~A.}\ \bibnamefont {Sobota}}, \bibinfo {author} {\bibfnamefont
  {S.}~\bibnamefont {Rebec}}, \bibinfo {author} {\bibfnamefont {J.~J.}\
  \bibnamefont {Lee}}, \bibinfo {author} {\bibfnamefont {T.}~\bibnamefont
  {Jia}}, \bibinfo {author} {\bibfnamefont {B.}~\bibnamefont {Moritz}},
  \bibinfo {author} {\bibfnamefont {C.}~\bibnamefont {Jia}}, \bibinfo {author}
  {\bibfnamefont {A.}~\bibnamefont {Gauthier}}, \bibinfo {author}
  {\bibfnamefont {Y.}~\bibnamefont {Li}}, \bibinfo {author} {\bibfnamefont
  {D.}~\bibnamefont {Leuenberger}}, \bibinfo {author} {\bibfnamefont
  {Y.}~\bibnamefont {Zhang}}, \bibinfo {author} {\bibfnamefont
  {L.}~\bibnamefont {Chaix}}, \bibinfo {author} {\bibfnamefont
  {W.}~\bibnamefont {Li}}, \bibinfo {author} {\bibfnamefont {H.}~\bibnamefont
  {Jang}}, \bibinfo {author} {\bibfnamefont {J.-S.}\ \bibnamefont {Lee}},
  \bibinfo {author} {\bibfnamefont {M.}~\bibnamefont {Yi}}, \bibinfo {author}
  {\bibfnamefont {G.~L.}\ \bibnamefont {Dakovski}}, \bibinfo {author}
  {\bibfnamefont {S.}~\bibnamefont {Song}}, \bibinfo {author} {\bibfnamefont
  {J.~M.}\ \bibnamefont {Glownia}}, \bibinfo {author} {\bibfnamefont
  {S.}~\bibnamefont {Nelson}}, \bibinfo {author} {\bibfnamefont {K.~W.}\
  \bibnamefont {Kim}}, \bibinfo {author} {\bibfnamefont {Y.-D.}\ \bibnamefont
  {Chuang}}, \bibinfo {author} {\bibfnamefont {Z.}~\bibnamefont {Hussain}},
  \bibinfo {author} {\bibfnamefont {R.~G.}\ \bibnamefont {Moore}}, \bibinfo
  {author} {\bibfnamefont {T.~P.}\ \bibnamefont {Devereaux}}, \bibinfo {author}
  {\bibfnamefont {W.-S.}\ \bibnamefont {Lee}}, \bibinfo {author} {\bibfnamefont
  {P.~S.}\ \bibnamefont {Kirchmann}}, \ and\ \bibinfo {author} {\bibfnamefont
  {Z.-X.}\ \bibnamefont {Shen}},\ }\href {\doibase 10.1126/science.aak9946}
  {\bibfield  {journal} {\bibinfo  {journal} {Science}\ }\textbf {\bibinfo
  {volume} {357}},\ \bibinfo {pages} {71} (\bibinfo {year} {2017})}\BibitemShut
  {NoStop}%
\bibitem [{\citenamefont {Perfetti}\ \emph {et~al.}(2018)\citenamefont
  {Perfetti}, \citenamefont {Loukakos}, \citenamefont {Lisowski}, \citenamefont
  {Bovensiepen}, \citenamefont {Wolf}, \citenamefont {Berger}, \citenamefont
  {Biermann},\ and\ \citenamefont {Geroges}}]{PerfettiTaS2}%
  \BibitemOpen
  \bibfield  {author} {\bibinfo {author} {\bibfnamefont {L.}~\bibnamefont
  {Perfetti}}, \bibinfo {author} {\bibfnamefont {P.}~\bibnamefont {Loukakos}},
  \bibinfo {author} {\bibfnamefont {M.}~\bibnamefont {Lisowski}}, \bibinfo
  {author} {\bibfnamefont {U.}~\bibnamefont {Bovensiepen}}, \bibinfo {author}
  {\bibfnamefont {M.}~\bibnamefont {Wolf}}, \bibinfo {author} {\bibfnamefont
  {H.}~\bibnamefont {Berger}}, \bibinfo {author} {\bibfnamefont
  {S.}~\bibnamefont {Biermann}}, \ and\ \bibinfo {author} {\bibfnamefont
  {A.}~\bibnamefont {Geroges}},\ }\href {\doibase
  10.1088/1367-2630/10/5/053019} {\bibfield  {journal} {\bibinfo  {journal}
  {New J. Phys}\ }\textbf {\bibinfo {volume} {10}},\ \bibinfo {pages} {053019}
  (\bibinfo {year} {2018})}\BibitemShut {NoStop}%
\bibitem [{\citenamefont {Zhu}\ \emph {et~al.}(2013)\citenamefont {Zhu},
  \citenamefont {Veenstra}, \citenamefont {Levy}, \citenamefont {Ubaldini},
  \citenamefont {Syers}, \citenamefont {Butch}, \citenamefont {Paglione},
  \citenamefont {Haverkort}, \citenamefont {Elfimov},\ and\ \citenamefont
  {Damascelli}}]{DamascelliBiSe2013}%
  \BibitemOpen
  \bibfield  {author} {\bibinfo {author} {\bibfnamefont {Z.-H.}\ \bibnamefont
  {Zhu}}, \bibinfo {author} {\bibfnamefont {C.~N.}\ \bibnamefont {Veenstra}},
  \bibinfo {author} {\bibfnamefont {G.}~\bibnamefont {Levy}}, \bibinfo {author}
  {\bibfnamefont {A.}~\bibnamefont {Ubaldini}}, \bibinfo {author}
  {\bibfnamefont {P.}~\bibnamefont {Syers}}, \bibinfo {author} {\bibfnamefont
  {N.~P.}\ \bibnamefont {Butch}}, \bibinfo {author} {\bibfnamefont
  {J.}~\bibnamefont {Paglione}}, \bibinfo {author} {\bibfnamefont {M.~W.}\
  \bibnamefont {Haverkort}}, \bibinfo {author} {\bibfnamefont {I.~S.}\
  \bibnamefont {Elfimov}}, \ and\ \bibinfo {author} {\bibfnamefont
  {A.}~\bibnamefont {Damascelli}},\ }\href {\doibase
  10.1103/PhysRevLett.110.216401} {\bibfield  {journal} {\bibinfo  {journal}
  {Phys. Rev. Lett.}\ }\textbf {\bibinfo {volume} {110}},\ \bibinfo {pages}
  {216401} (\bibinfo {year} {2013})}\BibitemShut {NoStop}%
\bibitem [{\citenamefont {Zhu}\ \emph {et~al.}(2014)\citenamefont {Zhu},
  \citenamefont {Veenstra}, \citenamefont {Zhdanovich}, \citenamefont
  {Schneider}, \citenamefont {Okuda}, \citenamefont {Miyamoto}, \citenamefont
  {Zhu}, \citenamefont {Namatame}, \citenamefont {Taniguchi}, \citenamefont
  {Haverkort}, \citenamefont {Elfimov},\ and\ \citenamefont
  {Damascelli}}]{DamascelliBiSe2014}%
  \BibitemOpen
  \bibfield  {author} {\bibinfo {author} {\bibfnamefont {Z.-H.}\ \bibnamefont
  {Zhu}}, \bibinfo {author} {\bibfnamefont {C.~N.}\ \bibnamefont {Veenstra}},
  \bibinfo {author} {\bibfnamefont {S.}~\bibnamefont {Zhdanovich}}, \bibinfo
  {author} {\bibfnamefont {M.~P.}\ \bibnamefont {Schneider}}, \bibinfo {author}
  {\bibfnamefont {T.}~\bibnamefont {Okuda}}, \bibinfo {author} {\bibfnamefont
  {K.}~\bibnamefont {Miyamoto}}, \bibinfo {author} {\bibfnamefont {S.-Y.}\
  \bibnamefont {Zhu}}, \bibinfo {author} {\bibfnamefont {H.}~\bibnamefont
  {Namatame}}, \bibinfo {author} {\bibfnamefont {M.}~\bibnamefont {Taniguchi}},
  \bibinfo {author} {\bibfnamefont {M.~W.}\ \bibnamefont {Haverkort}}, \bibinfo
  {author} {\bibfnamefont {I.~S.}\ \bibnamefont {Elfimov}}, \ and\ \bibinfo
  {author} {\bibfnamefont {A.}~\bibnamefont {Damascelli}},\ }\href {\doibase
  10.1103/PhysRevLett.112.076802} {\bibfield  {journal} {\bibinfo  {journal}
  {Phys. Rev. Lett.}\ }\textbf {\bibinfo {volume} {112}},\ \bibinfo {pages}
  {076802} (\bibinfo {year} {2014})}\BibitemShut {NoStop}%
\bibitem [{\citenamefont {Gierz}\ \emph {et~al.}(2011)\citenamefont {Gierz},
  \citenamefont {Henk}, \citenamefont {H\"ochst}, \citenamefont {Ast},\ and\
  \citenamefont {Kern}}]{Ast-graphene-interference}%
  \BibitemOpen
  \bibfield  {author} {\bibinfo {author} {\bibfnamefont {I.}~\bibnamefont
  {Gierz}}, \bibinfo {author} {\bibfnamefont {J.}~\bibnamefont {Henk}},
  \bibinfo {author} {\bibfnamefont {H.}~\bibnamefont {H\"ochst}}, \bibinfo
  {author} {\bibfnamefont {C.~R.}\ \bibnamefont {Ast}}, \ and\ \bibinfo
  {author} {\bibfnamefont {K.}~\bibnamefont {Kern}},\ }\href {\doibase
  10.1103/PhysRevB.83.121408} {\bibfield  {journal} {\bibinfo  {journal} {Phys.
  Rev. B}\ }\textbf {\bibinfo {volume} {83}},\ \bibinfo {pages} {121408}
  (\bibinfo {year} {2011})}\BibitemShut {NoStop}%
\bibitem [{\citenamefont {Liu}\ \emph {et~al.}(2011)\citenamefont {Liu},
  \citenamefont {Bian}, \citenamefont {Miller},\ and\ \citenamefont
  {Chiang}}]{Chiang-graphene}%
  \BibitemOpen
  \bibfield  {author} {\bibinfo {author} {\bibfnamefont {Y.}~\bibnamefont
  {Liu}}, \bibinfo {author} {\bibfnamefont {G.}~\bibnamefont {Bian}}, \bibinfo
  {author} {\bibfnamefont {T.}~\bibnamefont {Miller}}, \ and\ \bibinfo {author}
  {\bibfnamefont {T.-C.}\ \bibnamefont {Chiang}},\ }\href {\doibase
  10.1103/PhysRevLett.107.166803} {\bibfield  {journal} {\bibinfo  {journal}
  {Phys. Rev. Lett.}\ }\textbf {\bibinfo {volume} {107}},\ \bibinfo {pages}
  {166803} (\bibinfo {year} {2011})}\BibitemShut {NoStop}%
\bibitem [{\citenamefont {Kushwaha}\ \emph {et~al.}(2016)\citenamefont
  {Kushwaha}, \citenamefont {Pletikosic}, \citenamefont {Liang}, \citenamefont
  {Gyenis}, \citenamefont {Lapidus}, \citenamefont {Tian}, \citenamefont
  {Zhao}, \citenamefont {Burch}, \citenamefont {Jingjing}, \citenamefont
  {Wudi}, \citenamefont {Huiwen}, \citenamefont {Fedorov}, \citenamefont
  {Yazdani}, \citenamefont {Ong}, \citenamefont {Valla},\ and\ \citenamefont
  {Cava}}]{BSTSNatComm}%
  \BibitemOpen
  \bibfield  {author} {\bibinfo {author} {\bibfnamefont {S.~K.}\ \bibnamefont
  {Kushwaha}}, \bibinfo {author} {\bibfnamefont {I.}~\bibnamefont
  {Pletikosic}}, \bibinfo {author} {\bibfnamefont {T.}~\bibnamefont {Liang}},
  \bibinfo {author} {\bibfnamefont {A.}~\bibnamefont {Gyenis}}, \bibinfo
  {author} {\bibfnamefont {S.~H.}\ \bibnamefont {Lapidus}}, \bibinfo {author}
  {\bibfnamefont {Y.}~\bibnamefont {Tian}}, \bibinfo {author} {\bibfnamefont
  {H.}~\bibnamefont {Zhao}}, \bibinfo {author} {\bibfnamefont {K.~S.}\
  \bibnamefont {Burch}}, \bibinfo {author} {\bibfnamefont {L.}~\bibnamefont
  {Jingjing}}, \bibinfo {author} {\bibfnamefont {W.}~\bibnamefont {Wudi}},
  \bibinfo {author} {\bibfnamefont {J.}~\bibnamefont {Huiwen}}, \bibinfo
  {author} {\bibfnamefont {A.~V.}\ \bibnamefont {Fedorov}}, \bibinfo {author}
  {\bibfnamefont {A.}~\bibnamefont {Yazdani}}, \bibinfo {author} {\bibfnamefont
  {N.~P.}\ \bibnamefont {Ong}}, \bibinfo {author} {\bibfnamefont
  {T.}~\bibnamefont {Valla}}, \ and\ \bibinfo {author} {\bibfnamefont {R.~J.}\
  \bibnamefont {Cava}},\ }\href {\doibase 10.1038/ncomms11456} {\bibfield
  {journal} {\bibinfo  {journal} {Nat Commun}\ }\textbf {\bibinfo {volume}
  {7}},\ \bibinfo {pages} {11456} (\bibinfo {year} {2016})}\BibitemShut
  {NoStop}%
\bibitem [{\citenamefont {Kuroda}\ \emph {et~al.}(2016)\citenamefont {Kuroda},
  \citenamefont {Reimann}, \citenamefont {G\"udde},\ and\ \citenamefont
  {H\"ofer}}]{HoferTHZpumpTI}%
  \BibitemOpen
  \bibfield  {author} {\bibinfo {author} {\bibfnamefont {K.}~\bibnamefont
  {Kuroda}}, \bibinfo {author} {\bibfnamefont {J.}~\bibnamefont {Reimann}},
  \bibinfo {author} {\bibfnamefont {J.}~\bibnamefont {G\"udde}}, \ and\
  \bibinfo {author} {\bibfnamefont {U.}~\bibnamefont {H\"ofer}},\ }\href
  {\doibase 10.1103/PhysRevLett.116.076801} {\bibfield  {journal} {\bibinfo
  {journal} {Phys. Rev. Lett.}\ }\textbf {\bibinfo {volume} {116}},\ \bibinfo
  {pages} {076801} (\bibinfo {year} {2016})}\BibitemShut {NoStop}%
\bibitem [{\citenamefont {S\'anchez-Barriga}\ \emph {et~al.}(2014)\citenamefont
  {S\'anchez-Barriga}, \citenamefont {Scholz}, \citenamefont {Golias},
  \citenamefont {Rienks}, \citenamefont {Marchenko}, \citenamefont
  {Varykhalov}, \citenamefont {Yashina},\ and\ \citenamefont
  {Rader}}]{WarpingBiTeRader2014}%
  \BibitemOpen
  \bibfield  {author} {\bibinfo {author} {\bibfnamefont {J.}~\bibnamefont
  {S\'anchez-Barriga}}, \bibinfo {author} {\bibfnamefont {M.~R.}\ \bibnamefont
  {Scholz}}, \bibinfo {author} {\bibfnamefont {E.}~\bibnamefont {Golias}},
  \bibinfo {author} {\bibfnamefont {E.}~\bibnamefont {Rienks}}, \bibinfo
  {author} {\bibfnamefont {D.}~\bibnamefont {Marchenko}}, \bibinfo {author}
  {\bibfnamefont {A.}~\bibnamefont {Varykhalov}}, \bibinfo {author}
  {\bibfnamefont {L.~V.}\ \bibnamefont {Yashina}}, \ and\ \bibinfo {author}
  {\bibfnamefont {O.}~\bibnamefont {Rader}},\ }\href {\doibase
  10.1103/PhysRevB.90.195413} {\bibfield  {journal} {\bibinfo  {journal} {Phys.
  Rev. B}\ }\textbf {\bibinfo {volume} {90}},\ \bibinfo {pages} {195413}
  (\bibinfo {year} {2014})}\BibitemShut {NoStop}%
\bibitem [{\citenamefont {Sobota}\ \emph {et~al.}(2013)\citenamefont {Sobota},
  \citenamefont {Yang}, \citenamefont {Kemper}, \citenamefont {Lee},
  \citenamefont {Schmitt}, \citenamefont {Li}, \citenamefont {Moore},
  \citenamefont {Analytis}, \citenamefont {Fisher}, \citenamefont {Kirchmann},
  \citenamefont {Devereaux},\ and\ \citenamefont {Shen}}]{ShenBiSe2PPE}%
  \BibitemOpen
  \bibfield  {author} {\bibinfo {author} {\bibfnamefont {J.~A.}\ \bibnamefont
  {Sobota}}, \bibinfo {author} {\bibfnamefont {S.-L.}\ \bibnamefont {Yang}},
  \bibinfo {author} {\bibfnamefont {A.~F.}\ \bibnamefont {Kemper}}, \bibinfo
  {author} {\bibfnamefont {J.~J.}\ \bibnamefont {Lee}}, \bibinfo {author}
  {\bibfnamefont {F.~T.}\ \bibnamefont {Schmitt}}, \bibinfo {author}
  {\bibfnamefont {W.}~\bibnamefont {Li}}, \bibinfo {author} {\bibfnamefont
  {R.~G.}\ \bibnamefont {Moore}}, \bibinfo {author} {\bibfnamefont {J.~G.}\
  \bibnamefont {Analytis}}, \bibinfo {author} {\bibfnamefont {I.~R.}\
  \bibnamefont {Fisher}}, \bibinfo {author} {\bibfnamefont {P.~S.}\
  \bibnamefont {Kirchmann}}, \bibinfo {author} {\bibfnamefont {T.~P.}\
  \bibnamefont {Devereaux}}, \ and\ \bibinfo {author} {\bibfnamefont {Z.-X.}\
  \bibnamefont {Shen}},\ }\href {\doibase 10.1103/PhysRevLett.111.136802}
  {\bibfield  {journal} {\bibinfo  {journal} {Phys. Rev. Lett.}\ }\textbf
  {\bibinfo {volume} {111}},\ \bibinfo {pages} {136802} (\bibinfo {year}
  {2013})}\BibitemShut {NoStop}%
\bibitem [{\citenamefont {Cao}\ \emph {et~al.}(2013)\citenamefont {Cao},
  \citenamefont {Waugh}, \citenamefont {Zhang}, \citenamefont {Luo},
  \citenamefont {Wang}, \citenamefont {Reber}, \citenamefont {Mo},
  \citenamefont {Xu}, \citenamefont {Yang}, \citenamefont {Schneeloch},
  \citenamefont {Gu}, \citenamefont {Brahlek}, \citenamefont {Bansal},
  \citenamefont {Oh}, \citenamefont {Zunger},\ and\ \citenamefont
  {Dessau}}]{Cao2013}%
  \BibitemOpen
  \bibfield  {author} {\bibinfo {author} {\bibfnamefont {Y.}~\bibnamefont
  {Cao}}, \bibinfo {author} {\bibfnamefont {J.~A.}\ \bibnamefont {Waugh}},
  \bibinfo {author} {\bibfnamefont {X.-W.}\ \bibnamefont {Zhang}}, \bibinfo
  {author} {\bibfnamefont {J.-W.}\ \bibnamefont {Luo}}, \bibinfo {author}
  {\bibfnamefont {Q.}~\bibnamefont {Wang}}, \bibinfo {author} {\bibfnamefont
  {T.~J.}\ \bibnamefont {Reber}}, \bibinfo {author} {\bibfnamefont {S.~K.}\
  \bibnamefont {Mo}}, \bibinfo {author} {\bibfnamefont {Z.}~\bibnamefont {Xu}},
  \bibinfo {author} {\bibfnamefont {A.}~\bibnamefont {Yang}}, \bibinfo {author}
  {\bibfnamefont {J.}~\bibnamefont {Schneeloch}}, \bibinfo {author}
  {\bibfnamefont {G.~D.}\ \bibnamefont {Gu}}, \bibinfo {author} {\bibfnamefont
  {M.}~\bibnamefont {Brahlek}}, \bibinfo {author} {\bibfnamefont
  {N.}~\bibnamefont {Bansal}}, \bibinfo {author} {\bibfnamefont
  {S.}~\bibnamefont {Oh}}, \bibinfo {author} {\bibfnamefont {A.}~\bibnamefont
  {Zunger}}, \ and\ \bibinfo {author} {\bibfnamefont {D.~S.}\ \bibnamefont
  {Dessau}},\ }\href {\doibase 10.1038/nphys2685} {\bibfield  {journal}
  {\bibinfo  {journal} {Nat Phys}\ }\textbf {\bibinfo {volume} {9}},\ \bibinfo
  {pages} {499} (\bibinfo {year} {2013})}\BibitemShut {NoStop}%
\bibitem [{\citenamefont {Brida}\ \emph {et~al.}(2009)\citenamefont {Brida},
  \citenamefont {Bonora}, \citenamefont {Manzoni}, \citenamefont {Marangoni},
  \citenamefont {Villoresi}, \citenamefont {Silvestri},\ and\ \citenamefont
  {Cerullo}}]{BridaSetupMilan}%
  \BibitemOpen
  \bibfield  {author} {\bibinfo {author} {\bibfnamefont {D.}~\bibnamefont
  {Brida}}, \bibinfo {author} {\bibfnamefont {S.}~\bibnamefont {Bonora}},
  \bibinfo {author} {\bibfnamefont {C.}~\bibnamefont {Manzoni}}, \bibinfo
  {author} {\bibfnamefont {M.}~\bibnamefont {Marangoni}}, \bibinfo {author}
  {\bibfnamefont {P.}~\bibnamefont {Villoresi}}, \bibinfo {author}
  {\bibfnamefont {S.~D.}\ \bibnamefont {Silvestri}}, \ and\ \bibinfo {author}
  {\bibfnamefont {G.}~\bibnamefont {Cerullo}},\ }\href {\doibase
  10.1364/OE.17.012510} {\bibfield  {journal} {\bibinfo  {journal} {Opt.
  Express}\ }\textbf {\bibinfo {volume} {17}},\ \bibinfo {pages} {12510}
  (\bibinfo {year} {2009})}\BibitemShut {NoStop}%
\bibitem [{\citenamefont {Boschini}\ \emph {et~al.}(2015)\citenamefont
  {Boschini}, \citenamefont {Mansurova}, \citenamefont {Mussler}, \citenamefont
  {Kampmeier}, \citenamefont {Grützmacher}, \citenamefont {Braun},
  \citenamefont {Katmis}, \citenamefont {Moodera}, \citenamefont {Dallera},
  \citenamefont {Carpene}, \citenamefont {Franz}, \citenamefont {Czerner},
  \citenamefont {Heiliger}, \citenamefont {Kampfrath},\ and\ \citenamefont
  {Münzenberg}}]{BoschiniSciRep}%
  \BibitemOpen
  \bibfield  {author} {\bibinfo {author} {\bibfnamefont {F.}~\bibnamefont
  {Boschini}}, \bibinfo {author} {\bibfnamefont {M.}~\bibnamefont {Mansurova}},
  \bibinfo {author} {\bibfnamefont {G.}~\bibnamefont {Mussler}}, \bibinfo
  {author} {\bibfnamefont {J.}~\bibnamefont {Kampmeier}}, \bibinfo {author}
  {\bibfnamefont {D.}~\bibnamefont {Grützmacher}}, \bibinfo {author}
  {\bibfnamefont {L.}~\bibnamefont {Braun}}, \bibinfo {author} {\bibfnamefont
  {F.}~\bibnamefont {Katmis}}, \bibinfo {author} {\bibfnamefont {J.~S.}\
  \bibnamefont {Moodera}}, \bibinfo {author} {\bibfnamefont {C.}~\bibnamefont
  {Dallera}}, \bibinfo {author} {\bibfnamefont {E.}~\bibnamefont {Carpene}},
  \bibinfo {author} {\bibfnamefont {C.}~\bibnamefont {Franz}}, \bibinfo
  {author} {\bibfnamefont {M.}~\bibnamefont {Czerner}}, \bibinfo {author}
  {\bibfnamefont {C.}~\bibnamefont {Heiliger}}, \bibinfo {author}
  {\bibfnamefont {T.}~\bibnamefont {Kampfrath}}, \ and\ \bibinfo {author}
  {\bibfnamefont {M.}~\bibnamefont {Münzenberg}},\ }\href {\doibase
  10.1038/srep15304} {\bibfield  {journal} {\bibinfo  {journal} {Scientific
  Reports}\ }\textbf {\bibinfo {volume} {5}},\ \bibinfo {pages} {15304}
  (\bibinfo {year} {2015})}\BibitemShut {NoStop}%
\bibitem [{\citenamefont {Glinka}\ \emph {et~al.}(2015)\citenamefont {Glinka},
  \citenamefont {Babakiray}, \citenamefont {Johnson}, \citenamefont {Holcomb},\
  and\ \citenamefont {Lederman}}]{AcousticPhononBiSe}%
  \BibitemOpen
  \bibfield  {author} {\bibinfo {author} {\bibfnamefont {Y.~D.}\ \bibnamefont
  {Glinka}}, \bibinfo {author} {\bibfnamefont {S.}~\bibnamefont {Babakiray}},
  \bibinfo {author} {\bibfnamefont {T.~A.}\ \bibnamefont {Johnson}}, \bibinfo
  {author} {\bibfnamefont {M.~B.}\ \bibnamefont {Holcomb}}, \ and\ \bibinfo
  {author} {\bibfnamefont {D.}~\bibnamefont {Lederman}},\ }\href {\doibase
  10.1063/1.4919274} {\bibfield  {journal} {\bibinfo  {journal} {Journal of
  Applied Physics}\ }\textbf {\bibinfo {volume} {117}},\ \bibinfo {pages}
  {165703} (\bibinfo {year} {2015})}\BibitemShut {NoStop}%
\bibitem [{not()}]{noteReflectivity}%
  \BibitemOpen
  \href@noop {} {}\bibinfo {note} {Other probe photon energies show a similar
  behavior}\BibitemShut {NoStop}%
\bibitem [{\citenamefont {Bykov}\ \emph {et~al.}(2015)\citenamefont {Bykov},
  \citenamefont {Murzina}, \citenamefont {Olivier}, \citenamefont {Wurtz},\
  and\ \citenamefont {Zayats}}]{SHG_phonon_TI}%
  \BibitemOpen
  \bibfield  {author} {\bibinfo {author} {\bibfnamefont {A.~Y.}\ \bibnamefont
  {Bykov}}, \bibinfo {author} {\bibfnamefont {T.~V.}\ \bibnamefont {Murzina}},
  \bibinfo {author} {\bibfnamefont {N.}~\bibnamefont {Olivier}}, \bibinfo
  {author} {\bibfnamefont {G.~A.}\ \bibnamefont {Wurtz}}, \ and\ \bibinfo
  {author} {\bibfnamefont {A.~V.}\ \bibnamefont {Zayats}},\ }\href {\doibase
  10.1103/PhysRevB.92.064305} {\bibfield  {journal} {\bibinfo  {journal} {Phys.
  Rev. B}\ }\textbf {\bibinfo {volume} {92}},\ \bibinfo {pages} {064305}
  (\bibinfo {year} {2015})}\BibitemShut {NoStop}%
\bibitem [{\citenamefont {Sobota}\ \emph
  {et~al.}(2014{\natexlab{b}})\citenamefont {Sobota}, \citenamefont {Yang},
  \citenamefont {Leuenberger}, \citenamefont {Kemper}, \citenamefont
  {Analytis}, \citenamefont {Fisher}, \citenamefont {Kirchmann}, \citenamefont
  {Devereaux},\ and\ \citenamefont {Shen}}]{SobotaJElectr2014}%
  \BibitemOpen
  \bibfield  {author} {\bibinfo {author} {\bibfnamefont {J.}~\bibnamefont
  {Sobota}}, \bibinfo {author} {\bibfnamefont {S.-L.}\ \bibnamefont {Yang}},
  \bibinfo {author} {\bibfnamefont {D.}~\bibnamefont {Leuenberger}}, \bibinfo
  {author} {\bibfnamefont {A.}~\bibnamefont {Kemper}}, \bibinfo {author}
  {\bibfnamefont {J.}~\bibnamefont {Analytis}}, \bibinfo {author}
  {\bibfnamefont {I.}~\bibnamefont {Fisher}}, \bibinfo {author} {\bibfnamefont
  {P.}~\bibnamefont {Kirchmann}}, \bibinfo {author} {\bibfnamefont
  {T.}~\bibnamefont {Devereaux}}, \ and\ \bibinfo {author} {\bibfnamefont
  {Z.-X.}\ \bibnamefont {Shen}},\ }\href {\doibase
  http://dx.doi.org/10.1016/j.elspec.2014.01.005} {\bibfield  {journal}
  {\bibinfo  {journal} {Journal of Electron Spectroscopy and Related
  Phenomena}\ }\textbf {\bibinfo {volume} {195}},\ \bibinfo {pages} {249 }
  (\bibinfo {year} {2014}{\natexlab{b}})}\BibitemShut {NoStop}%
\bibitem [{\citenamefont {Hajlaoui}\ \emph {et~al.}(2014)\citenamefont
  {Hajlaoui}, \citenamefont {Papalazarou}, \citenamefont {Mauchain},
  \citenamefont {Perfetti}, \citenamefont {Taleb-Ibrahimi}, \citenamefont
  {Navarin}, \citenamefont {Monteverde}, \citenamefont {Auban-Senzier},
  \citenamefont {Pasquier}, \citenamefont {Moisan}, \citenamefont {Boschetto},
  \citenamefont {Neupane}, \citenamefont {Hasan}, \citenamefont {Durakiewicz},
  \citenamefont {Jiang}, \citenamefont {Xu}, \citenamefont {Miotkowski},
  \citenamefont {Chen}, \citenamefont {Jia}, \citenamefont {Ji}, \citenamefont
  {Cava},\ and\ \citenamefont {Marsi}}]{PerfettiBiTeNatComm}%
  \BibitemOpen
  \bibfield  {author} {\bibinfo {author} {\bibfnamefont {M.}~\bibnamefont
  {Hajlaoui}}, \bibinfo {author} {\bibfnamefont {E.}~\bibnamefont
  {Papalazarou}}, \bibinfo {author} {\bibfnamefont {J.}~\bibnamefont
  {Mauchain}}, \bibinfo {author} {\bibfnamefont {L.}~\bibnamefont {Perfetti}},
  \bibinfo {author} {\bibfnamefont {A.}~\bibnamefont {Taleb-Ibrahimi}},
  \bibinfo {author} {\bibfnamefont {F.}~\bibnamefont {Navarin}}, \bibinfo
  {author} {\bibfnamefont {M.}~\bibnamefont {Monteverde}}, \bibinfo {author}
  {\bibfnamefont {P.}~\bibnamefont {Auban-Senzier}}, \bibinfo {author}
  {\bibfnamefont {C.}~\bibnamefont {Pasquier}}, \bibinfo {author}
  {\bibfnamefont {N.}~\bibnamefont {Moisan}}, \bibinfo {author} {\bibfnamefont
  {D.}~\bibnamefont {Boschetto}}, \bibinfo {author} {\bibfnamefont
  {M.}~\bibnamefont {Neupane}}, \bibinfo {author} {\bibfnamefont
  {M.}~\bibnamefont {Hasan}}, \bibinfo {author} {\bibfnamefont
  {T.}~\bibnamefont {Durakiewicz}}, \bibinfo {author} {\bibfnamefont
  {Z.}~\bibnamefont {Jiang}}, \bibinfo {author} {\bibfnamefont
  {Y.}~\bibnamefont {Xu}}, \bibinfo {author} {\bibfnamefont {I.}~\bibnamefont
  {Miotkowski}}, \bibinfo {author} {\bibfnamefont {Y.}~\bibnamefont {Chen}},
  \bibinfo {author} {\bibfnamefont {S.}~\bibnamefont {Jia}}, \bibinfo {author}
  {\bibfnamefont {H.}~\bibnamefont {Ji}}, \bibinfo {author} {\bibfnamefont
  {R.}~\bibnamefont {Cava}}, \ and\ \bibinfo {author} {\bibfnamefont
  {M.}~\bibnamefont {Marsi}},\ }\href {\doibase 10.1038/ncomms4003} {\bibfield
  {journal} {\bibinfo  {journal} {Nat Comm}\ }\textbf {\bibinfo {volume} {5}},\
  \bibinfo {pages} {4003} (\bibinfo {year} {2014})}\BibitemShut {NoStop}%
\bibitem [{\citenamefont {Neupane}\ \emph {et~al.}(2015)\citenamefont
  {Neupane}, \citenamefont {Xu}, \citenamefont {Ishida}, \citenamefont {Jia},
  \citenamefont {Fregoso}, \citenamefont {Liu}, \citenamefont {Belopolski},
  \citenamefont {Bian}, \citenamefont {Alidoust}, \citenamefont {Durakiewicz},
  \citenamefont {Galitski}, \citenamefont {Shin}, \citenamefont {Cava},\ and\
  \citenamefont {Hasan}}]{TIgiganticlifetime_Hasan}%
  \BibitemOpen
  \bibfield  {author} {\bibinfo {author} {\bibfnamefont {M.}~\bibnamefont
  {Neupane}}, \bibinfo {author} {\bibfnamefont {S.-Y.}\ \bibnamefont {Xu}},
  \bibinfo {author} {\bibfnamefont {Y.}~\bibnamefont {Ishida}}, \bibinfo
  {author} {\bibfnamefont {S.}~\bibnamefont {Jia}}, \bibinfo {author}
  {\bibfnamefont {B.~M.}\ \bibnamefont {Fregoso}}, \bibinfo {author}
  {\bibfnamefont {C.}~\bibnamefont {Liu}}, \bibinfo {author} {\bibfnamefont
  {I.}~\bibnamefont {Belopolski}}, \bibinfo {author} {\bibfnamefont
  {G.}~\bibnamefont {Bian}}, \bibinfo {author} {\bibfnamefont {N.}~\bibnamefont
  {Alidoust}}, \bibinfo {author} {\bibfnamefont {T.}~\bibnamefont
  {Durakiewicz}}, \bibinfo {author} {\bibfnamefont {V.}~\bibnamefont
  {Galitski}}, \bibinfo {author} {\bibfnamefont {S.}~\bibnamefont {Shin}},
  \bibinfo {author} {\bibfnamefont {R.~J.}\ \bibnamefont {Cava}}, \ and\
  \bibinfo {author} {\bibfnamefont {M.~Z.}\ \bibnamefont {Hasan}},\ }\href
  {\doibase 10.1103/PhysRevLett.115.116801} {\bibfield  {journal} {\bibinfo
  {journal} {Phys. Rev. Lett.}\ }\textbf {\bibinfo {volume} {115}},\ \bibinfo
  {pages} {116801} (\bibinfo {year} {2015})}\BibitemShut {NoStop}%
\bibitem [{\citenamefont {S\'anchez-Barriga}\ \emph {et~al.}(2016)\citenamefont
  {S\'anchez-Barriga}, \citenamefont {Golias}, \citenamefont {Varykhalov},
  \citenamefont {Braun}, \citenamefont {Yashina}, \citenamefont {Schumann},
  \citenamefont {Min\'ar}, \citenamefont {Ebert}, \citenamefont {Kornilov},\
  and\ \citenamefont {Rader}}]{RaderTR_SPIN_ARPES}%
  \BibitemOpen
  \bibfield  {author} {\bibinfo {author} {\bibfnamefont {J.}~\bibnamefont
  {S\'anchez-Barriga}}, \bibinfo {author} {\bibfnamefont {E.}~\bibnamefont
  {Golias}}, \bibinfo {author} {\bibfnamefont {A.}~\bibnamefont {Varykhalov}},
  \bibinfo {author} {\bibfnamefont {J.}~\bibnamefont {Braun}}, \bibinfo
  {author} {\bibfnamefont {L.~V.}\ \bibnamefont {Yashina}}, \bibinfo {author}
  {\bibfnamefont {R.}~\bibnamefont {Schumann}}, \bibinfo {author}
  {\bibfnamefont {J.}~\bibnamefont {Min\'ar}}, \bibinfo {author} {\bibfnamefont
  {H.}~\bibnamefont {Ebert}}, \bibinfo {author} {\bibfnamefont
  {O.}~\bibnamefont {Kornilov}}, \ and\ \bibinfo {author} {\bibfnamefont
  {O.}~\bibnamefont {Rader}},\ }\href {\doibase 10.1103/PhysRevB.93.155426}
  {\bibfield  {journal} {\bibinfo  {journal} {Phys. Rev. B}\ }\textbf {\bibinfo
  {volume} {93}},\ \bibinfo {pages} {155426} (\bibinfo {year}
  {2016})}\BibitemShut {NoStop}%
\bibitem [{\citenamefont {Jozwiak}\ \emph {et~al.}(2016)\citenamefont
  {Jozwiak}, \citenamefont {Sobota}, \citenamefont {Gotlieb}, \citenamefont
  {Kemper}, \citenamefont {Rotundu}, \citenamefont {Birgeneau}, \citenamefont
  {Hussain}, \citenamefont {Lee}, \citenamefont {Shen},\ and\ \citenamefont
  {Lanzara}}]{Lanzara_SPIN_ARPES}%
  \BibitemOpen
  \bibfield  {author} {\bibinfo {author} {\bibfnamefont {C.}~\bibnamefont
  {Jozwiak}}, \bibinfo {author} {\bibfnamefont {J.~A.}\ \bibnamefont {Sobota}},
  \bibinfo {author} {\bibfnamefont {K.}~\bibnamefont {Gotlieb}}, \bibinfo
  {author} {\bibfnamefont {A.~F.}\ \bibnamefont {Kemper}}, \bibinfo {author}
  {\bibfnamefont {C.~R.}\ \bibnamefont {Rotundu}}, \bibinfo {author}
  {\bibfnamefont {R.~J.}\ \bibnamefont {Birgeneau}}, \bibinfo {author}
  {\bibfnamefont {Z.}~\bibnamefont {Hussain}}, \bibinfo {author} {\bibfnamefont
  {D.-H.}\ \bibnamefont {Lee}}, \bibinfo {author} {\bibfnamefont {Z.-X.}\
  \bibnamefont {Shen}}, \ and\ \bibinfo {author} {\bibfnamefont
  {A.}~\bibnamefont {Lanzara}},\ }\href {\doibase 10.1038/ncomms13143}
  {\bibfield  {journal} {\bibinfo  {journal} {Nature Communications}\ }\textbf
  {\bibinfo {volume} {7}},\ \bibinfo {pages} {13143} (\bibinfo {year}
  {2016})}\BibitemShut {NoStop}%
\bibitem [{\citenamefont {Richter}\ and\ \citenamefont
  {Becker}(1977)}]{Richter1977}%
  \BibitemOpen
  \bibfield  {author} {\bibinfo {author} {\bibfnamefont {W.}~\bibnamefont
  {Richter}}\ and\ \bibinfo {author} {\bibfnamefont {C.~R.}\ \bibnamefont
  {Becker}},\ }\href {\doibase 10.1002/pssb.2220840226} {\bibfield  {journal}
  {\bibinfo  {journal} {physica status solidi (b)}\ }\textbf {\bibinfo {volume}
  {84}},\ \bibinfo {pages} {619} (\bibinfo {year} {1977})}\BibitemShut
  {NoStop}%
\end{thebibliography}
\end{document}